\documentclass[aps,nofootinbib,showpacs,preprintnumbers,amsmath,amssymb]{revtex4}
\def\be{\begin{equation}}
\def\ee{\end{equation}}
\def\ba{\begin{eqnarray}}
\def\ea{\end{eqnarray}}
\def\bea{\begin{eqnarray}}
\def\eea{\end{eqnarray}}
\def\bes{\begin{subequations}}
\def\ees{\end{subequations}}
\newcommand{\A}{{\mathcal{A}}}
\newcommand{\tA}{{\widetilde {\mathcal{A}}}}
\newcommand{\ta}{{\widetilde a}}
\newcommand{\tlA}{{\mathfrak A}}
\newcommand{\td}{{\widetilde d}}
\newcommand{\tF}{{\widetilde F}}
\newcommand{\MSbar}{\overline{\rm MS}}  
\usepackage{graphics}
\usepackage{graphicx}
\usepackage{dcolumn}
\usepackage{bm}
\usepackage{epsfig}
\usepackage{graphicx,color}
\usepackage{multirow}

\begin{document}

\preprint{USM-TH-324}

\title{Perturbative QCD in acceptable schemes with holomorphic coupling\footnote{arXiv:1405.5815; v3: includes new Sec.~V - sum rule analysis for $\tau$ semihadronic decays in a proposed QCD scheme; new Refs.: [114,115,118-132]; to appear in Int.~J.~Mod.~Phys.~A}}

\author{Carlos Contreras$^1$}
\author{Gorazd Cveti\v{c}$^1$}
\author{Reinhart K\"ogerler$^3$}
\author{Pawel Kr\"oger$^2$}
\author{Oscar Orellana$^2$}
\affiliation{$^1$Department of Physics, Universidad T\'ecnica Federico Santa Mar\'ia, Casilla 110-V, Valpara\'iso, Chile\\
$^2$Department of Mathematics, Universidad T\'ecnica Federico Santa Mar\'ia, Casilla 110-V, Valpara\'iso, Chile\\
$^3$Department of Physics, Universit\"at Bielefeld, 33501 Bielefeld, Germany}
\date{\today}

\begin{abstract}
Perturbative QCD in mass independent schemes
leads in general to running coupling $a(Q^2)$ which is nonanalytic 
(nonholomorphic) in the regime of low spacelike momenta 
$|Q^2| \lesssim 1 \ {\rm GeV}^2$. Such (Landau) singularities are
inconvenient in the following sense: evaluations of spacelike physical
quantities ${\cal D}(Q^2)$ with such a running coupling $a(\kappa Q^2)$ 
($\kappa \sim 1$) give us expressions with the
same kind of singularities, while the general principles of local 
quantum field theory require that the mentioned physical quantities
have no such singularities. In a previous work, certain classes
of perturbative mass independent beta functions were found such that 
the resulting coupling was holomorphic. However, the resulting
perturbation series showed explosive increase of coefficients already
at ${\rm N}^4{\rm LO}$ order, as a consequence of the requirement that
the theory reproduce the correct value of the $\tau$ lepton 
semihadronic strangeless decay ratio $r_{\tau}$.
In this work we successfully extend
the construction to specific classes of perturbative 
beta functions such that the perturbation
series do not show explosive increase of coefficients, the 
perturbative coupling is holomorphic, and the correct value of $r_{\tau}$
is reproduced. In addition, we extract, with Borel sum rule analysis of 
the $V+A$ channel of the semihadronic strangeless decays of $\tau$ lepton,
reasonable values of the corresponding $D=4$ and $D=6$ condensates.
\end{abstract}

\maketitle

\section{Introduction}
\label{sec:intr}

The perturbative QCD (pQCD) calculations are 
usually performed in mass independent
schemes, i.e., schemes in which
beta function $\beta(a)$ of the running coupling 
(couplant) $a(Q^2)$ ($\equiv \alpha_s(Q^2)/\pi$) 
has expansion in powers of $a$ such that the
beta expansion coefficients
depend on the number of effective quark flavors $N_f$. When the squared
momenta $q^2 \equiv -Q^2$ are low, $|Q^2| \lesssim 1 \ {\rm GeV}^2$,
the mentioned coefficients have $N_f=3$.
Such calculations give for the running coupling $a(Q^2)$ a function
which has, for the general spacelike momenta
$Q^2 \in \mathbb{C} \backslash (-\infty, 0]$, 
nonholomorphic (singular) behavior in the small momentum regime
$|Q^2| \lesssim 1 \ {\rm GeV}^2$, and these singularities are
usually called Landau ghosts (or Landau singularities). 
When any spacelike physical
quantities ${\cal D}(Q^2)$, such as the current correlators and 
structure functions, are evaluated in pQCD as (truncated) series
involving such coupling $a(\kappa Q^2)$ (where $\kappa \sim 1$ is
a positive renormalization scale parameter), the resulting expressions 
${\cal D}(Q^2)_{\rm eval.} = {\cal F}(a(\kappa Q^2))$ manifest the same type
of singularities for $Q^2 \in \mathbb{C} \backslash (-\infty, 0]$
(and $|Q^2| \lesssim 1 \ {\rm GeV}^2$). Such singularities
of ${\cal D}(Q^2)_{\rm eval.}$ are physically unacceptable, because
${\cal D}(Q^2)$ must be an analytic (holomorphic) function of $Q^2$
in the entire complex $Q^2$ plane with the exception of 
the negative semiaxis $Q^2 \in \mathbb{C} \backslash (-\infty, -M_{\rm thr}^2]$
(where the threshold mass is $M_{\rm thr} \sim 0.1$ GeV), this
being a consequence of general principles of (local) quantum
field theories \cite{BS,Oehme}. 
Even resummations of infinite number of terms in the perturbation
expansion of ${\cal D}(Q^2)$, practicable in QCD for example in the
large-$\beta_0$ approximation, do not cure the problem of Landau
ghosts [cf.~comments following Eqs.~(\ref{dLB}) in Appendix A]. 
If we are to apply a universal running
coupling $a(Q^2)$ in the evaluation of a low-momentum quantity ${\cal D}(Q^2)$,
the analytic properties of $a(Q^2)$ should reflect the mentioned analytic
properties of ${\cal D}(Q^2)$.

The notion of a universal running coupling $a(Q^2)$ is intimately connected with the concept of perturbation expansion. Since perturbation theory is directly applicable only to those physical quantities or, to those circumstances (momenta, etc.), which are characterized by small coupling, originally only such coupling makes direct sense. Within QCD this is the coupling in the regime of high momenta (asymptotic freedom) where partons (quarks and gluons) do exist in the usual sense. Nevertheless, one can attribute a meaning to a universal running coupling outside the high momentum regime. One of the preconditions for the applicability of such a coupling is that the aforementioned nonanalyticity (Landau singularities) of $a(Q^2)$, at low $|Q^2| \lesssim 1 \ {\rm GeV}^2$ in the complex $Q^2$ plane outside the negative semiaxis, does not appear or is eliminated.

A formalism exists which extends the use of the universal running
coupling to the regime $|Q| \sim 1$ GeV, namely the Operator Product
Expansion (OPE) in the sense of the ITEP School (pQCD+OPE), 
Refs.~\cite{Shifman1,Shifman2}.
In such approach the inclusive spacelike quantities ${\cal D}(Q^2)$, 
are evaluated by adding to the usual perturbation expansion of the
term with the lowest dimension (leading-twist term),
${\cal F}(a(\kappa Q^2))$, other terms 
which involve vacuum expectation values (condensates) of various operators
$O_N$ with higher dimensions $2 N$, i.e., terms proportional to
$\langle O_N \rangle/Q^{2N}$.
Complete formalism which
would extend the regime of applicability of this pQCD+OPE approach
at present does not exist, but attempts have
been made in this direction with the use of nonlocal condensates 
\cite{NLC}.

Various independent lines of research support the existence of
the concept of the running coupling $a(Q^2)$ in low-momentum regime and
suggest that it is finite and possibly holomorphic there: the
Gribov-Zwanziger approach \cite{Gr1,Gr2,Gr3,Gr4}; calculations involving 
Dyson-Schwinger equations (DSE) for gluon and ghost propagators 
and vertices 
\cite{DSE1a,DSE1b,DSE1c,DSE1d,DSE1e,DSE2a,DSE2b,DSE2c,DSE2d,DSE2e,DSE2f,AHS}; 
stochastic quantization \cite{STQ};
functional renormalization group equations \cite{FRG1,FRG2,FRG3};
lattice calculations \cite{latt1a,latt1b,latt1c,latt1d}.
In addition, the finiteness of the coupling at $Q \to 0$ is
suggested by specific applications \cite{StMatt1,StMatt2} of
the Principle of Minimal Sensitivity (PMS) \cite{Stevenson,St2a,St2b},
by models using the AdS/CFT
correspondence modified by a dilaton backgound \cite{AdS1,AdS2},
in scenarios with larger quark flavor number $N_f$ \cite{SteveIRFP1,SteveIRFP2},
and in various other approaches such
as those in Refs.~\cite{Simonov1,Simonov2,BKSKKSh1,BKSKKSh2,BKSKKSh3,BKSKKSh4,BKSKKSh5,Deur,Court}.

The nonanalyticity of $a(Q^2)$ in low-momentum 
regimes in the usual pQCD schemes was addressed in the seminal works of
Shirkov, Solovtsov {\it et al.\/} \cite{ShS1,ShS2,MS,Sh1Sh2a,Sh1Sh2b}, 
where a holomorphic
version $\A^{\rm (APT)}(Q^2)$
of the pQCD coupling $a(Q^2)$ (in any  mass independent 
scheme) was constructed, via a use of
the Cauchy theorem and dispersion integral in which the offending
(Landau) cut of $a(Q^2)$ was eliminated and the cut of $a(Q^2)$ for
$Q^2 < 0$ was left unchanged; in a sense, this is a ``minimal'' analytization
approach, widely referred in the literature as Analytic Perturbation
Theory (APT). This approach includes the analogous construction,
via dispersive integral, of the holomorphic analogs  $\A_n^{\rm (APT)}(Q^2)$ of
the (integer) powers $a(Q^2)^n$ of pQCD coupling. The formalism was later
extended to the construction of APT analogs of any physical
quantity \cite{KS}, and of APT analogs $\A_{\nu}^{\rm (FAPT)}(Q^2)$ of
noninteger powers $a(Q^2)^{\nu}$ ($\nu$ noninteger) in the
works \cite{BMS1,BMS2,BMS3} (Fractional Analytic Perturbation
Theory - FAPT).  For a
review of FAPT, see Refs.~\cite{Bakulev1,Bakulev2}, and mathematical packages 
for numerical calculation are given in Refs.~\cite{BK1,BK2,BK3}.

Since the publication of APT \cite{ShS1,ShS2,MS},
several other (extended) analytic QCD models, i.e., models
of holomorphic $\A(Q^2)$, have been constructed 
\cite{Nest1a,Nest1b,Nest1c,Nest1d,Nest2a,Nest2b,Nest2c,Webber,CV12a,CV12b,Alekseev,1danQCD,2danQCD,Shirkovmass}.\footnote{
In Refs.~\cite{Nest1a,Nest1b,Nest1c,Nest1d} the
coupling is holomorphic for $Q^2 \in \mathbb{C} \backslash (-\infty, 0]$
and is infinite at $Q^2=0$.} 
Analytic QCD models [(F)APT and others] and related dispersive
approaches have been used in various contexts 
\cite{MSS1,MSS2a,MSS2b,MagrGl,mes2a,mes2b,DeRafael,MagrTau1,MagrTau2,Nest3a,Nest3b,Nest3c,anOPE,anOPE2a,anOPE2b}.
For reviews of some analytic QCD models, see Refs.~\cite{Prosperi,Shirkov}.

Furthermore, the higher power analogs
 $\A_{\nu}(Q^2)$ of $a(Q^2)^{\nu}$ in such general analytic QCD models
are constructed by the procedures of Refs.~\cite{CV12a,CV12b} (when $\nu$ is integer)
and Ref.~\cite{GCAK} (when $\nu$ is general real). 

It turns out that all these holomorphic couplings $\A(Q^2)$ are nonperturbative,
i.e., for $|Q^2| > \Lambda^2$ (where $\Lambda^2 \sim 1 \ {\rm GeV}^2$)
they differ from the corresponding pQCD coupling $a(Q^2)$ (i.e., $a(Q^2)$
in the same scheme) by terms $\A(Q^2) - a(Q^2) \sim (\Lambda^2/Q^2)^N$,
where $N=1$ in the models of
Refs.~\cite{ShS1,ShS2,Nest1a,Nest1b,Nest1c,Nest1d,Nest2a,Nest2b,Nest2c,CV12a,CV12b,Shirkovmass},
$N=3,4,5$ in the models of Refs.~\cite{1danQCD,Alekseev}, \cite{Webber}, 
and \cite{2danQCD}, respectively. However, the power terms
$\Lambda^2/Q^2$, at high $|Q^2|$ (small $a(Q^2)$), can be expressed as
$\exp[-1/( \beta_0 a(Q^2))] \sim e^{-1/a}$, which is a nonanalytic function
in $a$ (around $a=0$). This implies that the analytic QCD models
cannot be described by a perturbative beta function $\beta(a) \equiv
d a(Q^2)/d \ln Q^2$, i.e., by a $\beta(a)$ function which is described
at small $|a|$ fully by its Taylor expansion in powers of $a$. The function
$\beta(\A)$ in all these analytic QCD models contains terms
$\sim e^{-1/\A}$.

In this context, the following  question appears naturally: does there
exist a perturbative $\beta(a)$ function [$\beta(a) = - \beta_0 a^2
- \beta_1 a^3 - \beta_2 a^4 - \ldots$] such that the corresponding
(perturbative) running coupling $a(Q^2)$ is a holomorphic function
in the complex plane $Q^2 \in \mathbb{C} \backslash (-\infty, -M_{\rm thr}^2]$
(or  $Q^2 \in \mathbb{C} \backslash (-\infty, 0]$)?
In Refs.~\cite{anpQCD1,anpQCD2}, an extensive attempt was made to obtain such
an analytic pQCD (anpQCD). The major obstacles to such an effort turned out to
be the simultaneous fulfillment of two requirements: a) $a(Q^2)$
is holomorphic; b) the value of the best measured low-energy QCD observable
$r_{\tau} = 0.203 \pm 0.004$ can be reproduced in this anpQCD. Here,
$r_{\tau}$ is the QCD massless canonical part [$r_{\tau} = a + {\cal O}(a^2)$] 
of the $(V+A)$-channel of the $\tau$ lepton strangeless semihadronic decay ratio $R_{\tau}(\Delta S=0)$, 
and in $r_{\tau}$ the quark mass effects 
have been subtracted and the chirality-conserving higher-twist
effects are known to be very suppressed \cite{DDHMZ}. The two requirements 
a) and b) have the
tendency to be mutually exclusive: almost any anpQCD gives far too
low value ($< 0.14$) of $r_{\tau}$; if the free parameters in the considered
classes of perturbative $\beta(a)$ functions are varied in such a way that the
value $0.203$ of $r_{\tau}$ is approached (from below), the coupling
$a(Q^2)$ in general acquires singularities inside the plane
$Q^2 \in \mathbb{C} \backslash (-\infty, 0]$ and thus ceases to be
holomorphic. The problem of too low value of $r_{\tau}$ was already
encountered earlier \cite{MSS2a,MSS2b,Geshkenbein}
for the analytic (and nonperturbative) QCD model APT 
of Refs.~\cite{ShS1,ShS2,MS,Sh1Sh2a,Sh1Sh2b}.
Nonetheless, in Refs.~\cite{anpQCD1,anpQCD2}, for specific classes
of perturbative $\beta(a)$ functions with holomorphic $a(Q^2)$, 
the $\beta(a)$ functions were modified/multiplied by another 
perturbative function
$f_{\rm fact}(a)$  such that the
perturbation expansion of $r_{\tau}$, including its first four (known) terms,
gave the correct value $0.203$ and the analyticity of $a(Q^2)$ was 
preserved. However, the price to pay was that the resulting beta function 
acquired in its expansion very large $\beta_4$ coefficient at $a^5$
($\beta_4 \sim 10^6$-$10^7$) and thus the fifth term $\sim a^5$ in the
expansion of $r_{\tau}$ became uncontrollably high.

In this work we return to this problem and find an attractive solution
to the mentioned problem, by constructing such perturbative
${\cal F}_{\rm fact}(a)$ functions that give
perturbative beta functions $\beta(a) \propto {\cal F}_{\rm fact}(a)$ 
that simultaneously: 
(a) keep the perturbation expansion coefficients under control to an
arbitrarily high order; (b) reproduce the correct value
$r_{\tau}=0.203$; (c) preserve the analyticity of $a(Q^2)$ in
$Q^2 \in \mathbb{C} \backslash (-\infty, -M_{\rm thr}^2]$.
In Sec.~\ref{sec:form} we present the formalism of integration
of the renormalization group equation in the complex $Q^2$ plane and various
conditions (analyticity, universality) that have to to be fulfilled.
In Sec.~\ref{sec:nofact} we reproduce several classes of $\beta$ functions 
that give holomorphic $a(Q^2)$ but fail to achieve the value
of $r_{\tau}=0.203$. In Sec.~\ref{sec:fact} we introduce the
functions ${\cal F}_{\rm fact}(a)$ with which we modify/multiply the
beta functions of the previous Section and which give us the 
acceptable perturbative analytic QCD framework: holomorphic
$a(Q^2)$, the correct value $r_{\tau}=0.203$, and the perturbation
expansion coefficients under control. 
In Sec.~\ref{sec:BSR} we perform, 
with one of the obtained perturbative analytic QCD schemes,
an analysis with Borel sum rules of the $V+A$ channel of the semihadronic
decays of $\tau$ lepton and extract reasonable values of the
corresponding $D=4$ and $D=6$ condensates.
In Sec.~\ref{sec:summ} we summarize our results.

\section{Conditions, integration}
\label{sec:form}

The running coupling $a(Q^2) \equiv \alpha_s(Q^2)/\pi$ in QCD fulfills the 
renormalization group equation (RGE)
\be
Q^2 \frac{d a(Q^2)}{d Q^2} = \beta \left( a(Q^2) \right) \ ,
\label{RGE1}
\ee
where $\beta(a)$ is beta function. In the approach of the
construction of the perturbative and holomorphic coupling
$a(Q^2)=\A(Q^2)$ here,
the starting point will be the construction of beta function
$\beta(a)$, and then the coupling function $a(Q^2)$ will be obtained
by numerical integration of the RGE (\ref{RGE1}) in
the complex $Q^2$ plane.
We will impose three central requirements on
$\beta(a)$ and the resulting $a(Q^2)$ functions:
\begin{enumerate}
\item
The coupling $a(Q^2)$ is a perturbative (pQCD coupling);
this is equivalent to the requirement that beta function is a 
holomorphic (analytic) function of $a$ at $a=0$ 
\be
\beta(a) = - \beta_0 a^2 - \beta_1 a^3 - \beta_2 a^4 - \dots =
 - \beta_0 a^2 (1 + c_1 a+ c_2 a^2 + \ldots ) \ ,
\label{betaexp}
\ee
cf.~Refs.~\cite{anpQCD1,anpQCD2,Raczka1,Raczka2,Raczka3}. 
For example, $\beta(a)$ cannot contain
the typically nonperturbative terms $\sim \exp[-C/a(Q^2)]$ for which
the Taylor expansion around $a=0$ is blind.
\item
The coupling $a(Q^2)$ must reproduce the correct 
measured value $r_{\tau} = 0.203 \pm 0.004$,
where $r_{\tau}$ is the QCD massless canonical part 
[$r_{\tau} = a + {\cal O}(a^2)$] 
of $\tau$ lepton strangeless semihadronic decay ratio $R_{\tau}(\Delta S=0)$ 
(with the quark mass effects subtracted and the higher-twist
effects suppressed). We recall that $r_{\tau}$ is 
at the moment the best measured inclusive low-energy QCD observable.
\item
The coupling $a(Q^2)$,
constructed by the integration of the RGE (\ref{RGE1}),
must be a holomorphic function, i.e., holomorphic in the
complex $Q^2$ plane $Q^2 \in \mathbb{C} \backslash (-\infty, -M_{\rm thr}^2]$,
where the threshold mass is $M_{\rm thr} \sim 0.1$ GeV.
\end{enumerate}

In the integration of Eq.~(\ref{RGE1}), we first need the
initial condition $a(Q^2_{\rm in})$ at an initial (low) scale
$Q^2_{\rm in}$. Since we are interested in the holomorphic behavior
of $a(Q^2)$ at not very high $|Q^2|$ ($|Q^2| \lesssim (3 m_c)^2$), 
we consider for simplicity the heavy quarks $c, b, t$ to be decoupled,
and the three light quarks $u,d,s$ we will consider to be massless.
Stated otherwise, the number of active flavors
in the RGE (\ref{RGE1}) is $N_f=3$. We choose our initial scale to be 
$Q^2_{\rm in} = (3 m_c)^2 \approx 14.52 \ {\rm GeV}^2$.
In order to obtain the value of $a(Q^2_{\rm in})$, i.e., in the
scheme determined by the considered beta function $\beta(a)$,
we should first obtain the value ${\overline a}(Q^2_{\rm in})$ in
the $\MSbar$ scheme. For this, we take the present world average \cite{PDG2012}
${\overline a}(M_Z^2)=0.1184/\pi$ and RGE-run it down
by the known 4-loop $\MSbar$ beta function ${\overline \beta}(a)$
from $Q^2=M_Z^2$ to $Q^2_{\rm in} = (3 m_c)^2$. The quark thresholds
are taken at $Q_{\rm thr} = \kappa {\overline m}_b$ and 
$Q_{\rm thr} = \kappa {\overline m}_c$, 
according to the 3-loop matching conditions
\cite{CKS}, where $1 \leq \kappa \sim 1$ and
we take $\kappa=3$, and the $\MSbar$ masses
${\overline m}_b=4.20$ GeV and ${\overline m}_c=1.27$ GeV.
This gives us ${\overline a}(Q^2_{\rm in}) \approx 0.0716$ (at $N_f=3$).
The corresponding value of $a(Q^2_{\rm in})$ is then obtained by
using the integrated form of RGE (i.e., implicit solution) in its subtracted
form, cf.~Appendix A of Ref.~\cite{Stevenson}
(cf.~also Appendix A of Ref.~\cite{CK})
\ba
\lefteqn{
\frac{1}{a} + c_1 \ln \left( \frac{c_1 a}{1\!+\!c_1 a} \right)
+ \int_0^a dx \left[ \frac{ \beta(x) + \beta_0 x^2 (1\!+\!c_1 x)}
{x^2 (1\!+\!c_1 x) \beta(x) } \right]}
\nonumber\\
&& =
\frac{1}{{\overline a}} + c_1 \ln \left( \frac{c_1 {\overline a}}{1\!+\!c_1 {\overline a}} \right)
+ \int_0^{\overline a} dx \left[ \frac{ {\overline \beta}(x) + \beta_0 x^2 (1\!+\!c_1 x)}
{x^2 (1\!+\!c_1 x) {\overline \beta}(x) } \right],
\label{match}
\ea
where $a \equiv a(Q^2_{\rm in}) = a_{\rm in}$ and
${\overline a} \equiv {\overline a}(Q^2_{\rm in}) \approx 0.0716$.
We note that $\beta_0= (1/4)(11 - 2 N_f/3)$ and 
$c_1 = \beta_1/\beta_0 =(1/4) (102-38 N_f/3)/(11-2 N_f/3)$ 
are the universal beta-function coefficients in the 
mass independent schemes
($\beta_0=9/4$ and $c_1=16/9$ for $N_f=3$), while the other
expansion coefficients 
$c_j \equiv \beta_j/\beta_0$ ($j \geq 2$) in Eq.~(\ref{betaexp})
characterize the scheme \cite{Stevenson}. Any choice of
$\beta$ function then determines, via Eq.~(\ref{match}), the
initial value $a(Q^2_{\rm in}) = a_{\rm in}$ for the numerical
integration of Eq.~(\ref{RGE1}).\footnote{
Eq.~(3) states that the left-hand side is exactly independent of the
renormalization scheme parameters $c_j$ ($j \geq 2$) appearing in
the expansion (\ref{betaexp}) of $\beta(x)$; this (exact) independence 
was proven in Appendix A of Ref.~\cite{Stevenson}.}

\begin{figure}[htb] 
\centering\includegraphics[width=140mm]{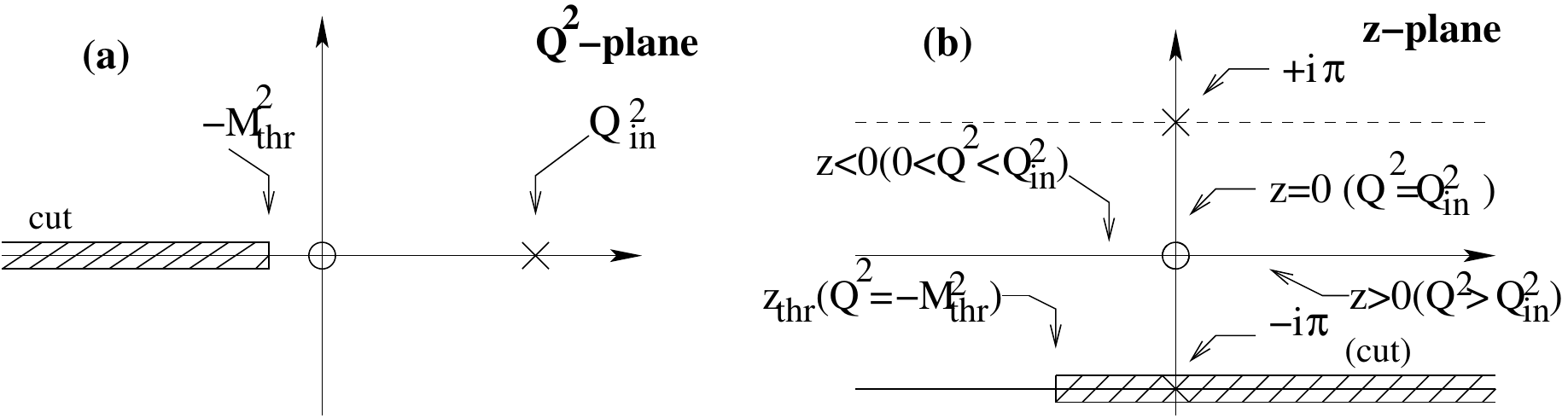}
 \caption{\footnotesize  (a) Complex $Q^2$ plane; (b) complex
$z$ plane where $z=\ln(Q^2/Q^2_{\rm in})$; the non-timelike
stripe is $-\pi \leq {\rm Im}z < + \pi$.}
\label{Qzfig}
 \end{figure}

As mentioned, the RGE (\ref{RGE1}) will be solved numerically not just for $Q^2>0$,
but in the entire complex $Q^2$ plane.
Following the presentation in Refs.~\cite{anpQCD1,anpQCD2},
a new complex variable is introduced: 
$z = \ln(Q^2/Q^2_{\rm in})$.
Then the first sheet of the complex $Q^2$ plane corresponds to
the semiopen stripe $- \pi \leq {\rm Im}z < + \pi$ in the complex $z$ plane.
The general spacelike regime $Q^2 \in \mathbb{C} \backslash (-\infty,0]$,
where $a(Q^2)$ is holomorphic in the considered perturbative
analytic QCD (anpQCD) scenarios, is represented by the open
stripe $- \pi < {\rm Im}(z) < + \pi$ in the $z$-plane. The timelike
(Minkowskian) semiaxis $Q^2 \leq 0$ corresponds to the border line
${\rm Im}z = - \pi$ of the $z$-stripe.
The point $Q^2=0$ is $z=-\infty$, and $Q^2=Q^2_{\rm in}$
($\approx 14.52 \ {\rm GeV}^2$) is $z=0$. In
Figs.~\ref{Qzfig} (a) and (b) we present the corresponding
general spacelike and timelike regimes in the complex $Q^2$ plane and
on the $z$ stripe, respectively, with a view that
$a(Q^2)$ is holomorphic in the extended regime
$Q^2 \in \mathbb{C} \backslash (-\infty, -M_{\rm thr}^2]$
(where $0 \leq M_{\rm thr} \lesssim 0.1$ GeV).
Let us denote $a(Q^2) \equiv F(z)$. Then RGE (\ref{RGE1}) acquires the form
\be
\frac{d F(z)}{d z} = \beta(F(z)) 
\qquad \left( F(z) \equiv a(Q^2) \right)
\label{RGEz}
\ee
in terms of $z = \ln(Q^2/Q^2_{\rm in})$ in the semiopen stripe 
$- \pi \leq {\rm Im}z < + \pi$. The requirement that
$a(Q^2)$ be holomorphic for $Q^2 \in \mathbb{C} \backslash (-\infty,0]$
now means that $F(z)$ is holomorphic ($ \Rightarrow
\partial F/\partial {\overline z} = 0$) in the open stripe 
$- \pi < {\rm Im}(z) < + \pi$. The (physical) singularities can appear
only on the timelike line ${\rm Im} (z) = - \pi$.
Let us denote $z=x+i y$ and $F=u + i v$;
then we can rewrite RGE (\ref{RGEz}) as a coupled system of real
partial differential equations for $u(x,y)$ and $v(x,y)$
\bes
\label{RGExy}
\ba
\frac{\partial u(x,y)}{\partial x} &=& {\rm Re} \beta (u + i v) \ ,
\quad 
\frac{\partial v(x,y)}{\partial x} = {\rm Im} \beta (u + i v) \ ,
\label{RGEx}
\\
\frac{\partial u(x,y)}{\partial y} &=& - {\rm Im} \beta (u + i v) \ ,
\quad 
\frac{\partial v(x,y)}{\partial y} = {\rm Re} \beta (u + i v) \ .
\label{RGEy}
\ea
\ees
We recall that $x=\ln(|Q^2|/Q^2_{\rm in})$, 
$y={\rm arg}(Q^2)$ ($-\pi \leq y < \pi$), 
$u={\rm Re} \; a$, $v={\rm Im}\; a$.
Having chosen an Ansatz for beta function $\beta(a(Q^2)) \equiv \beta(F(z))$ 
and the corresponding initial condition value $a(Q^2_{\rm in})$, 
the integration of equations (\ref{RGExy}) is then implemented numerically
to high precision by the MATHEMATICA software \cite{Math}.
Numerical analyses suggest that it is very difficult to obtain in this
way analytic coupling $a(Q^2)$ for 
$Q^2 \in \mathbb{C} \backslash (-\infty,0]$, i.e., 
analytic $F(z)$ in the entire open stripe
$-\pi < {\rm Im} z < \pi$, unless we require in addition also analyticity
in and around the point $Q^2=0$ ($z = -\infty$).
Stated otherwise, with certain classes of pQCD $\beta$-functions
we obtain the correct holomorphic behavior of $a(Q^2)$.
We represent the analyticity in $Q^2$ at $Q^2=0$ in the form
\be
a(Q^2) = a_0 + a_1 (Q^2/\Lambda^2) + {\cal O}[ (Q^2/\Lambda^2)^2] \ ,
\label{aQ0}
\ee
where $a_0 = a(Q^2=0)=F(z=-\infty) < \infty$, and $a_1 \not=0$. 
Applying the derivative $d/dz = d/d \ln Q^2$ to
this series, the condition reads
\be
\beta^{\prime}(F) \vert_{F=a_0} = +1 \ .
\label{anQ0b}
\ee
The Ans\"atze for beta function are thus taken in the form 
(cf.~Refs.~\cite{anpQCD1,anpQCD2})
 \be
\beta(F) = - \beta_0 F^2 (1 - Y) f(Y) \vert_{Y \equiv F/a_0} \ ,
\label{ans1}
\ee
where the function $f(Y)$ fulfills three conditions
\bes
\label{con}
\ba
f(Y) \ && {\rm analytic \ at} \ Y=0  \qquad {\rm (pQCD)} \ ,
\label{conpQCD}
\\
f(Y) & = & 1 + (1 + c_1 a_0) Y + {\cal O}(Y^2)  \qquad {\rm (pQCD)} \ ,
\label{conc1}
\\
a_0 \beta_0 f(1) &=& 1 \qquad (Q^2=0 \ {\rm analyticity}) \ .
\label{conanQ0}
\ea
\ees
The first condition says that the beta function is perturbative; the second accounts for the universality of the $c_1$ coefficient of the pQCD
expansion (\ref{betaexp}); the third condition is the aforementioned
condition of analyticity of $a(Q^2)$ at $Q^2=0$, i.e., Eq.~(\ref{aQ0}).
Let us once more recall that the choice of a specific type of 
beta function corresponds to a specific renormalization scheme,
characterized by the coefficients $c_j$ ($j \geq 2$)
of the series expansion (\ref{betaexp}) of this beta function.

If $a_1=0$ in Eq.~(\ref{aQ0}) and $f(Y)$ is any rational (Pad\'e)
function, Landau singularities appear, as argued in Ref.~\cite{anpQCD2}
(footnote 3 and Appendix A there); e.g., when 
$a(Q^2) = a_0 + {\cal O}((Q^2/\Lambda^2)^n)$ with $n \geq 2$,
Landau poles of $F(z)$ appear at ${\rm Im} \; z = \pm \pi/n$.

The condition of analyticity at $Q^2=0$, i.e.~Eq.~(\ref{aQ0}),
implies that there is a finite region of analyticity of $a(Q^2)$ around
$Q^2=0$, i.e., that the branching point of the cut
$(-\infty, -M_{\rm thr}^2]$ of $a(Q^2)$ in the complex $Q^2$ plane starts at
a nonzero threshold energy $-M_{\rm thr}^2 < 0$. This implicitly signals that
the masses of light pseudoscalar mesons $\pi$ and $K$ are nonzero,
i.e., that the masses of $u$, $d$ and $s$ quarks are not strictly zero.
Therefore, the condition (\ref{aQ0}) implicitly incorporates these
effects, which would otherwise be very difficult to incorporate explicitly
with nonzero light quark masses in the RGE. Another, more practical,
reason for imposing the condition (\ref{aQ0}) 
lies in the fact that it turned out to be 
very difficult or impossible to achieve numerically analyticity
of $a(Q^2)$ in the Euclidean complex plane 
$Q^2  \in \mathbb{C} \backslash (-\infty, 0]$ unless the point
$Q^2=0$ was also included as a point of analyticity of $a(Q^2)$,
Refs.~\cite{anpQCD1,anpQCD2}.

Often in pQCD, the PMS \cite{Stevenson,St2a,St2b}
and effective charge (ECH) \cite{ECH1,ECH2,ECH3,KKP,Gupta1,Gupta2} schemes
(at $n$-loop level, $n$ finite) are constructed from 
a truncated perturbation series
${\cal D}_{\rm pt}(Q^2)^{[n]}$ (i.e., including the terms up to $a^n$) of
a considered spacelike observable ${\cal D}(Q^2)$ in such a way
that, in the PMS procedure all the terms $\sim a^{n+2}$ 
are consistently discarded in the derivatives 
$\partial{\cal D}(Q^2)^{[n]}_{\rm PMS}/\partial {\rm RS} =0$
(where RS=$\ln \mu^2, c_2, c_3, \ldots$), 
and in the ECH procedure all the terms $\sim a^{n+1}$ 
are consistently discarded in 
${\cal D}(Q^2)^{[n]}_{\rm ECH}$ (for example, cf.~Refs.~\cite{KataevStarshenko}). 
Such schemes have scheme coefficients $c_j$ ($j=2,\ldots,n-1$) 
which are independent of $Q^2$
of the considered observable ${\cal D}(Q^2)$. If such (PMS or ECH)
schemes give finite $a_0 \equiv a(0)$, e.g. those with $c_2<0$, they in 
general do not result in holomorphic coupling $a(Q^2)$, at least not 
at $Q^2=0$, because $a_0$ in general does not fulfill the condition 
(\ref{conanQ0}). If $a_0 \beta_0 f(1) > 1$, there are Landau singularities
and poles inside the $z$ stripe (cf.~Appendix A of Ref.~\cite{anpQCD2}); 
if  $a_0 \beta_0 f(1) < 1$, it can happen
that no singularities appear inside the $z$ stripe and the only point of 
nonanalyticity is $Q^2=0$; but then the value of $r_{\tau}$ is even generally
much more below the experimental value $r_{\tau} =0.203$,
Ref.~\cite{progress}.

\section{Beta functions and results}
\label{sec:nofact}

\begin{figure}[htb] 
\begin{minipage}[b]{.49\linewidth}
\centering\includegraphics[width=85mm]{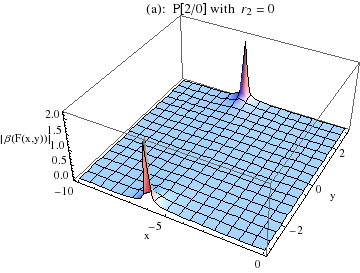}
\end{minipage}
\begin{minipage}[b]{.49\linewidth}
\centering\includegraphics[width=85mm]{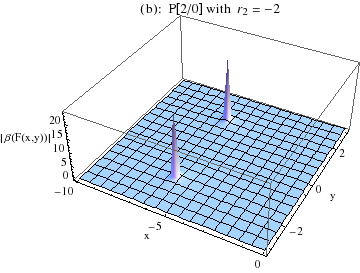}
\end{minipage}
\vspace{-0.2cm}
 \caption{\footnotesize  $|\beta(F(z))|$ as a function of
$z=x+i y$ for the beta function (\ref{ans1}) with $f(Y)$
having the form (\ref{fP20}) with: (a) $r_2=0$;
(b) $r_2=-2$.}
\label{figP20bt}
 \end{figure}

\begin{figure}[htb] 
\begin{minipage}[b]{.49\linewidth}
\centering\includegraphics[width=85mm]{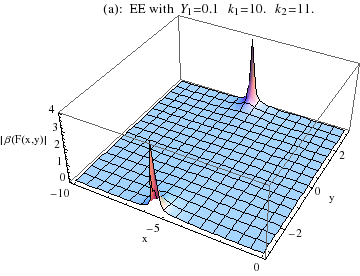}
\end{minipage}
\begin{minipage}[b]{.49\linewidth}
\centering\includegraphics[width=85mm]{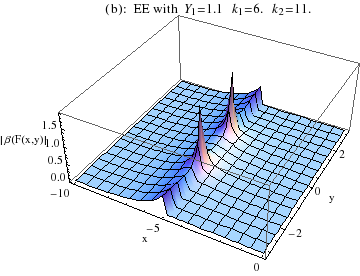}
\end{minipage}
\vspace{-0.2cm}
 \caption{\footnotesize  (a) $|\beta(F(z))|$ as a function of
$z=x+i y$, where $\beta$ has the form (\ref{ans1}) with $f(Y)$
having the ``EE'' form (\ref{EE}) 
with the values of free parameters $Y_1, k_1,k_2$
as indicated;
(b) same as in (a), but with different values of parameters $Y_1$ and $k_1$.}
\label{figEEbt}
 \end{figure}

\begin{figure}[htb] 
\begin{minipage}[b]{.49\linewidth}
\centering\includegraphics[width=85mm]{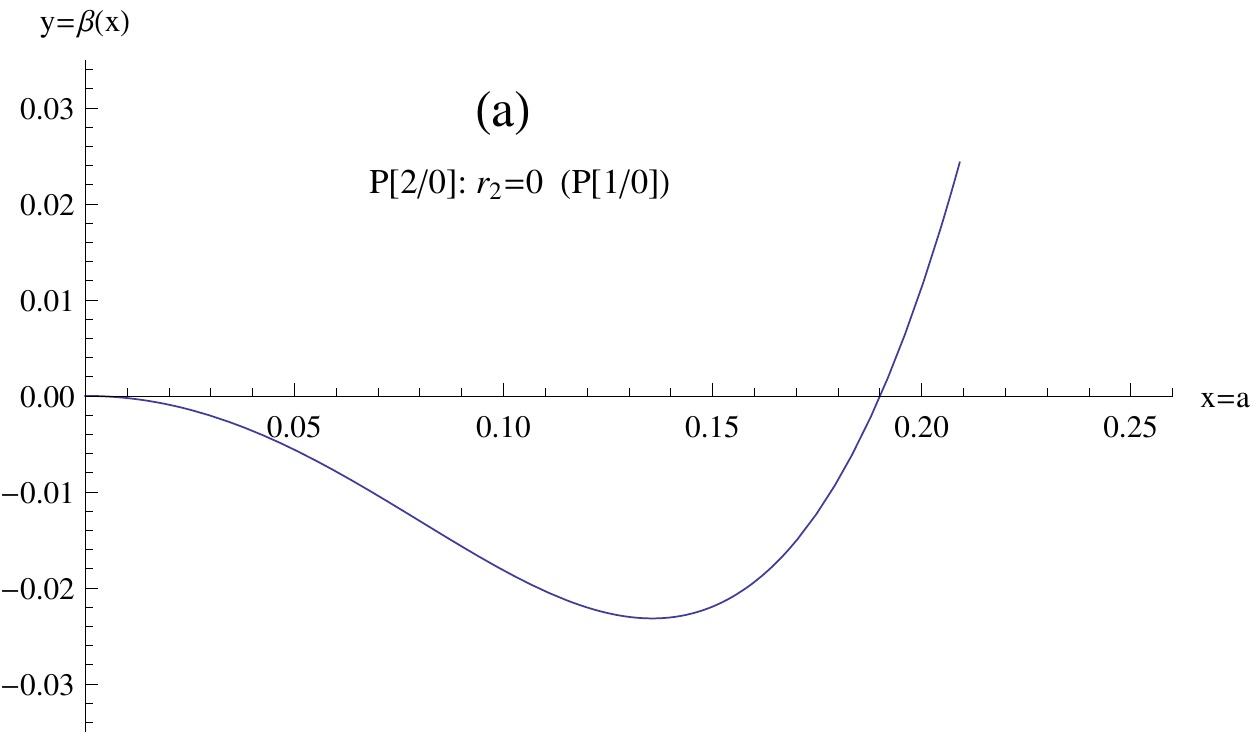}
\end{minipage}
\begin{minipage}[b]{.49\linewidth}
\centering\includegraphics[width=85mm]{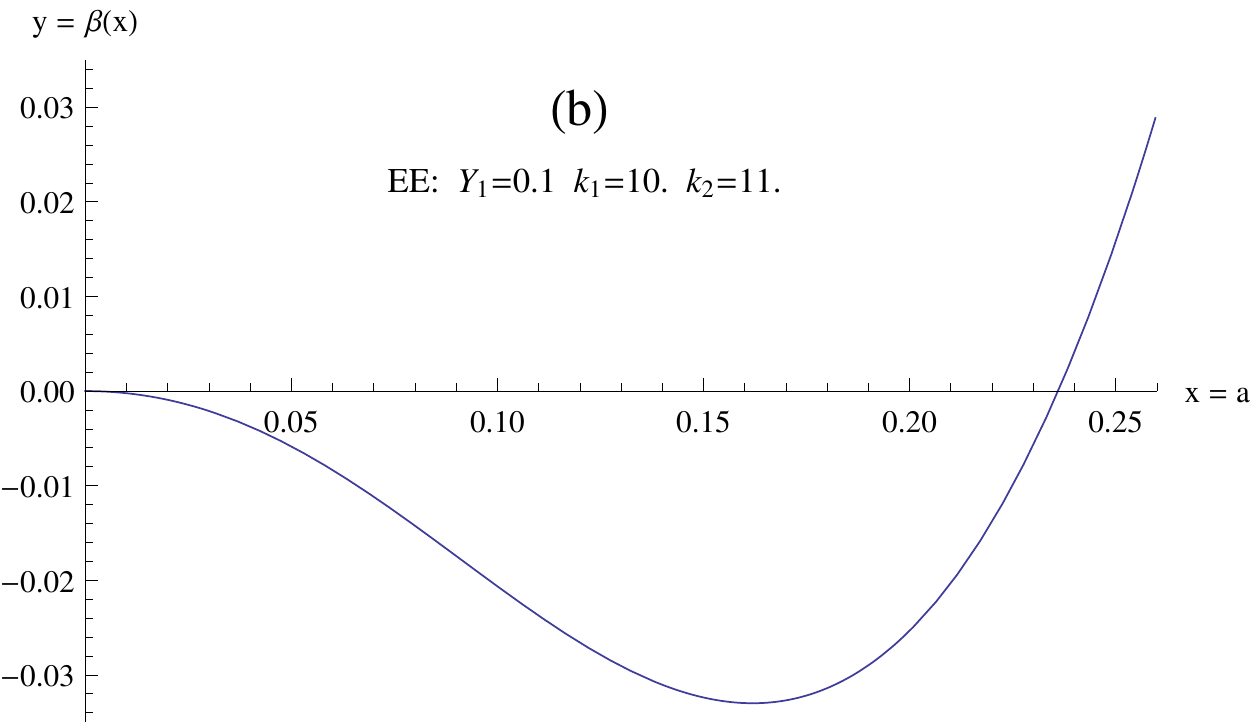}
\end{minipage}
\vspace{-0.2cm}
 \caption{\footnotesize  Beta function
$\beta(a)$ as a function of positive $a$: (a) for the
case when $f(Y)$ has the form (\ref{fP20}) with $r_2=0$,
i.e., linear polynomial;
(b) for the case when $f(Y)$ is the exponential-related ``EE'' function
(\ref{EE}) with $Y_1=0.1; k_1=10; k_2=11$.}
\label{figbet}
 \end{figure}

\begin{figure}[htb] 
\begin{minipage}[b]{.49\linewidth}
\centering\includegraphics[width=85mm]{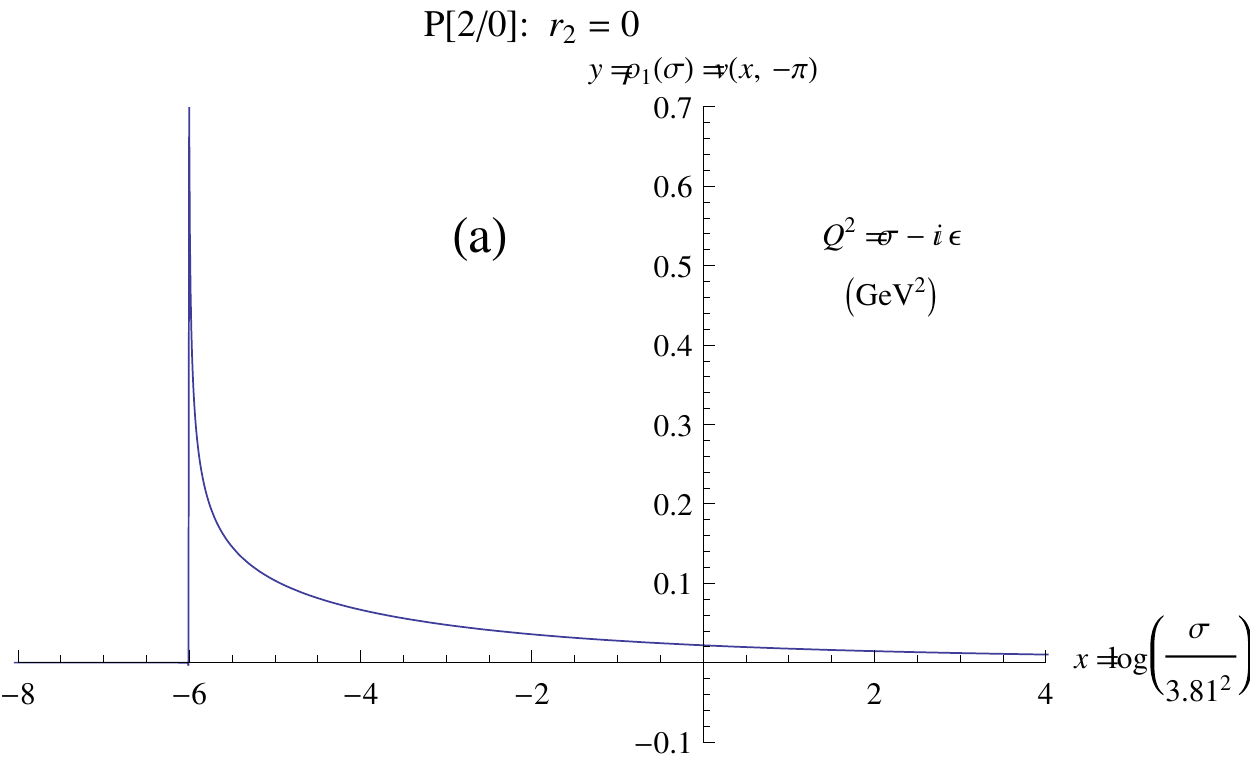}
\end{minipage}
\begin{minipage}[b]{.49\linewidth}
\centering\includegraphics[width=85mm]{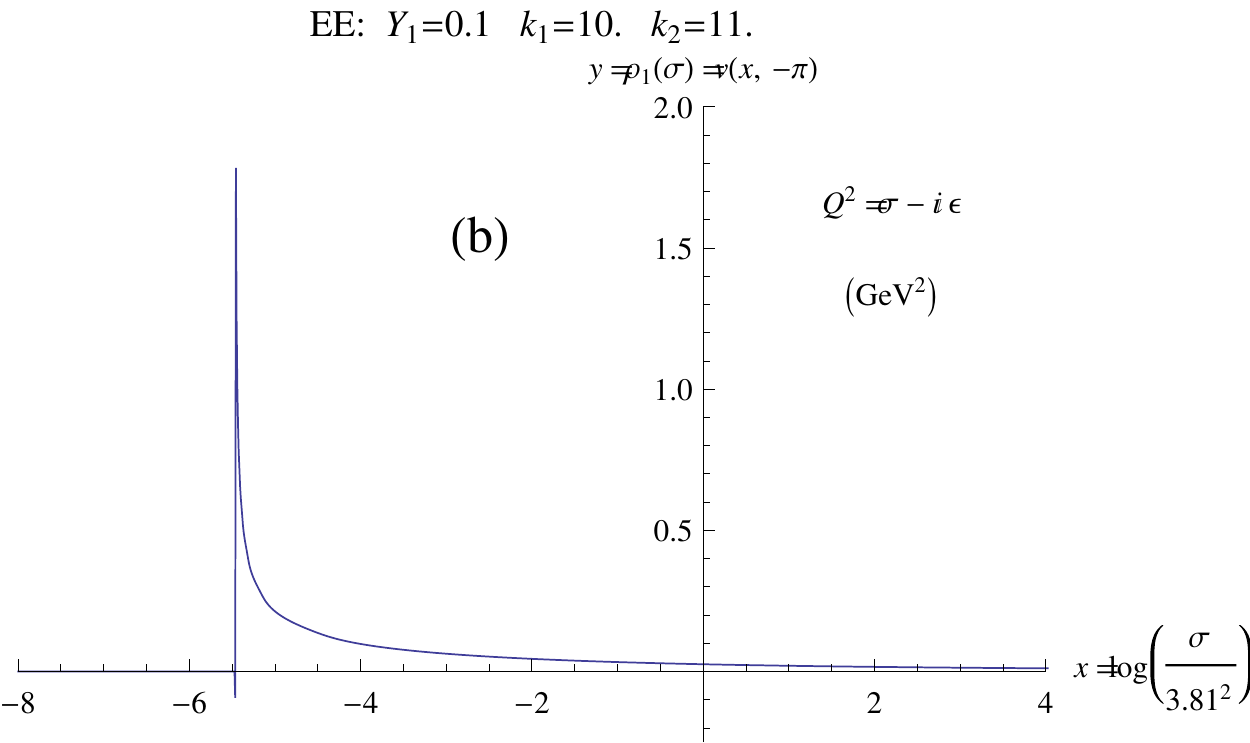}
\end{minipage}
\vspace{-0.2cm}
 \caption{\footnotesize  The discontinuity function
$\rho_1(\sigma) = {\rm Im} \ a(Q^2=-\sigma - i \epsilon) =
{\rm Im} F(z=x - i \pi) = v(x,y=-\pi)$ as a function of
$x = {\rm Re}(z) = \ln(\sigma/Q^2_{\rm in})$, for the two cases
(\ref{fP20}) and (\ref{EE}).}
\label{figrho}
 \end{figure}

Among $f(Y)$ functions that satisfy the three conditions (\ref{con}),
only certain specific subsets, with free parameters within $f(Y)$ varying
in restricted intervals, lead upon the numerical
integration of RGE's (\ref{RGExy}) to holomorphic behavior,
i.e., to a holomorphic $F(z)$ in the entire open stripe
 $-\pi < {\rm Im} z < \pi$. However, the evaluation of the
aforementioned $\tau$ lepton decay ratio $r_{\tau}$ gave
us consistently values well below the experimental values
$ 0.203 \pm 0.004$, namely values below $0.15$.
In Refs.~\cite{anpQCD1,anpQCD2}, for representation of the numerical results,
various Ans\"atze were used for the function $f(Y)$:
(1) in the form of polynomials; (2) Pad\'e's (ratios of polynomials);
(3) product of rescaled and translated functions of the type
$(e^{-Y}-1)/Y$ and $Y/(e^{-Y}-1)$, respectively. As mentioned,
it turned out that, while such functions did give us holomorphic $F(z)$ 
in the entire open stripe of $z$, they gave for $r_{\tau}$ far too low values
($<0.15$). Here we summarize some of the results of Refs.~\cite{anpQCD1,anpQCD2}
for the three mentioned cases.
\begin{enumerate}
\item 
The case of quadratic polynomial $f(Y)$ 
\be
f(Y) = 1 + r_1 Y + r_2 Y^2 \ ,
\label{fP20}
\ee
where the first coefficient is $r_1=(1 + c_1 a_0)$ due to the
condition (\ref{conc1}). In order to see whether the resulting
running coupling $a(Q^2)=F(z) \equiv F(x+i y)$ has or has not
singularities within the physical stripe $-\pi < z < \pi$
(Landau singularities), we present in Figs.~\ref{figP20bt}(a) and (b) the
results for the quantity $|\beta(F(z))|$ which should manifest singularities 
at the same $z$ values as the singularities of $F(z)$. 
In the case $r_2=0$, Fig.~\ref{figP20bt}(a) 
suggests that there are no Landau singularities, i.e.,
no singularities on the open stripe $- \pi < {\rm Im} z < \pi$,
only singularitires on the timelike axis (${\rm Im} z = \pm \pi$).
In the case $r_2 <0$ ($r_2=-2$ taken), 
Fig.~\ref{figP20bt}(b) clearly shows that
there are Landau singularities. As argued in Ref.~\cite{anpQCD2}, 
for $0 \leq r_2 < r_1^2/4$  there are no Landau poles. In the $r_2=0$ case
there are no free parameters, because the apparently free parameters 
$r_1$ and $a_0=a(Q^2=0)$ 
are fixed by the conditions (\ref{conc1})-(\ref{conanQ0}):
$a_0=0.1901$ and $r_1=(1 + c_1 a_0)=1.338$. We did not choose $r_2>0$
because, although the coupling is holomorphic, 
the resulting $r_{\tau}$ is even lower than in the
$r_2=0$ case. We refer to the case
(\ref{fP20}) with $r_2=0$ as P[1/0] because $F(Y)$ is 
Pad\'e ${\rm P}[1/0](Y)$ in this case.\footnote{
The model of Ref.~\cite{StMatt1,StMatt2}, based on the 
principle of minimal sensitivity (PMS) \cite{Stevenson,St2a,St2b}, has the
same form of beta function, with the conditions (\ref{conpQCD})-(\ref{conc1})
fulfilled, but $a_0 \equiv a(0)$ does not satisfy the condition
(\ref{conanQ0}), which in this case states: $a_0 \beta_0 (2 + c_1 a_0)=1$.
Namely, the model of Ref.~\cite{StMatt1,StMatt2} has 
$a_0 \beta_0 (2 + c_1 a_0)>1$, the coupling is thus not analytic at
$Q^2=0$. In the version of the PMS approach applied in 
Ref.~\cite{StMatt1,StMatt2},
the resulting scheme coefficients depend on the squared momentum $Q^2_0$
of the considered observable ${\cal D}(Q_0^2)$.
It is possible that the coupling is analytic in the rest of the
$Q^2$-plane (except on the semiaxis $(-\infty,0]$) when this approach
is applied to a considered observable ${\cal D}(Q_0^2)$ carefully
at each (complex) $Q_0^2$  value. 
On the other hand, when it is applied to ${\cal D}(Q_0^2)$ at a fixed
chosen $Q_0^2$, the resulting PMS coupling $a(Q^2)$ in general has Landau 
singularities inside the $Q^2$ plane.}
\item
The case of Pad\'e [1/1] $f(Y)$
\be
f(Y) = \frac{(1 - t_1 Y)}{(1 - u_1 Y)} \ .
\label{fP11}
\ee
We have seemingly three parameters ($t_1$, $u_1$ and $a_0$), but two of them
are eliminated by the conditions (\ref{conc1})-(\ref{conanQ0}). We can
regard as the only free parameter the coefficient $u_1$. It turns out
that for $u_1 =-0.1$ we obtain approximately largest $r_{\tau}$ while
still no Landau poles.
\item
The case of $f(Y)$ being  a product of rescaled and translated 
functions of the type $(e^{-Y}-1)/Y$ and $Y/(e^{-Y}-1)$
\be
{\rm EE:} \; 
f(Y) =  \frac{ \left( \exp[- k_1 (Y - Y_1)] -1 \right) }
{ [ k_1 (Y - Y_1) ] }
\frac{ [ k_2 (Y - Y_2) ] }{ \left( \exp[- k_2 (Y - Y_2)] -1 \right) }
\times {\cal K}(k_1,Y_1,k_2,Y_2) .
\label{EE}
\ee 
Here, the constant ${\cal K}$ ensures the required normalization
$f(Y=0)=1$. At first sight, we have five free parameters:
$a_0 \equiv a(Q^2=0)$ and
four parameters for translation and rescaling
($Y_1$, $k_1$, $Y_2$, $k_2$). Two of the parameters
($Y_2$ and $a_0$) are eliminated by the conditions 
(\ref{conc1})-(\ref{conanQ0}). Further, $0 < k_1 < k_2$ must be
fulfilled to get physically acceptable behavior. 
Figs.~\ref{figEEbt}(a), (b) represent the numerical results
for $|\beta(F(z))|$ for the following two chosen cases:
(a) $Y_1=0.1; k_1=10; k_2=11$ ($\Rightarrow Y_2 \approx 0.1839$);
(b) $Y_1=1.1; k_1=6; k_2=11$ ($\Rightarrow Y_2 \approx 0.2386$). 
Figs.~\ref{figEEbt} suggest that the case (a) has no Landau
singularities, and that the case (b) clearly has Landau singularities.
The case EE(a) is such that $a(Q^2)$ is kept holomorphic and
simultaneously the value of $r_{\tau}$ is higher than for most
of other choices of EE parameters (but still not high enough, see later).
\end{enumerate}
In Figs.~\ref{figbet}(a), (b)  we present beta function $\beta(a)$
as a function of (positive) $a$,
and in Figs.~\ref{figrho}(a), (b) the discontinuity function
$\rho_1(\sigma) \equiv {\rm Im} a(-\sigma-i \epsilon)$ as a function of 
$\ln(\sigma/Q^2_{\rm in})$, for two cases of holomorphic $a(Q^2)$:
for the case P[1/0] (i.e., P[2/0] with $r_2=0$), and for
the case EE ($Y_1=0.1; k_1=10; k_2=11$), respectively.
We note that the discontinuity function in both cases shows a clear
sudden threshold jump, i.e., the cut starts at $Q^2 = - M^2_{\rm thr}$,
where $M_{\rm thr}  =0.189$ GeV and $0.248$ GeV, respectively.

Nonetheless, the $r_{\tau}$ values, which are calculated as a
leading-$\beta_0$ (LB) resummation plus beyond-the-leading-$\beta_0$
terms (bLB: NLB $\sim \ta_2 \sim a^2$, ${\rm N}^2{\rm LB} \sim \ta_3 \sim a^3$, 
${\rm N}^3{\rm LB} \sim \ta_4 \sim a^4$),\footnote{
We refer to Appendix \ref{app1} and Refs.~\cite{anpQCD1,anpQCD2} for details of
calculation of $r_{\tau}$; and to Appendix \ref{app2} for the evaluation
of the expansion coefficients in the perturbation expansion of the
underlying spacelike quantity in the general renormalization schemes.}
are far too low
in the described cases, see Table \ref{tabrtau1}. We recall that
the free parameters in the three aforementioned cases (P[1/0], P[1/1], EE)
were chosen in such a way as to make $r_{\tau}$ as big as possible
while simultaneously preserving the holomorphic property
of $a(Q^2)$. The EE Ansatz gives us the highest $r_{\tau} \approx 0.145$,
but still a long way from the experimental value $r_{\tau} =  0.203 \pm 0.004$.
\begin{table}
\caption{The four terms (and their sum) in truncated analytic expansions
for $r_{\tau}$, with LB-contributions resummed and the three bLB
terms organized in contour integrals of $\ta_{n+1}$ 
[cf.~Eqs.~(\ref{rtman2})-(\ref{TnIta}), with Eq.~(\ref{LBrt2})
and the definition (\ref{tan})]. 
RScl parameter is $\kappa=1$ (i.e., $\mu^2=Q^2$
on the contour). Included are beta function coefficients $c_j$ ($j=2,3,4$),
$a_{\rm in} \equiv a((3 m_c)^2)$ and
$a_0 \equiv a(0)$ values, and the threshold mass value $M_{\rm thr}$ (in GeV).}
\label{tabrtau1}  
\begin{ruledtabular}
\begin{tabular}{l|lllllccclll}
$f$ & $r_{\tau}:$ LB  & NLB & ${\rm N}^2{\rm LB}$ & ${\rm N}^3{\rm LB}$ & sum & 
$c_2$ & $c_3$ & $c_4$ & $a_{\rm in}$ & $a_0$  &  $M_{\rm thr}$ [GeV]
\\
\hline
P[1/0] & 0.1122 & 0.0006 & 0.0137& 0.0007 & 0.1272 & -37.02 & 0  & 0 & 0.0600 & 0.1901 & 0.189 
\\
\hline
P[1/1] & 0.1130 & 0.0006 & 0.0144 & 0.0005 & 0.1285 & -37.54 &  18.84 & -9.46 & 0.0600 & 0.1992 & 0.179
\\
\hline
EE  & 0.1364 & 0.0009 & 0.0025 & 0.0048 & 0.1445 & -10.80 & -157.62 & -644.32 & 0.0649 & 0.2360 & 0.248
\end{tabular}
\end{ruledtabular}
\end{table}
In the Table \ref{tabrtau1} we also display the first few scheme
coefficients $c_j$ ($j=2,3,4$), the value of $a_0 \equiv a(0)$, and
the threshold mass $M_{\rm thr}$ (in GeV).\footnote{
These values differ slightly from the corresponding values in Tables II and III of Ref.~\cite{anpQCD2}, because here we use for the RGE-running from $Q^2=M_Z^2$ to $Q^2=Q^2_{\rm in}$ ($=(3 m_c)^2$) the 4-loop truncated 
(polynomial) beta $\MSbar$
function (in Refs.~\cite{anpQCD1,anpQCD2} it was the corresponding Pad\'e P[2/3]$(a)$
function), and now we use the world average value $\alpha_s(M_Z^2,\MSbar)=0.1184$ \cite {PDG2012} (in Refs.~\cite{anpQCD1,anpQCD2} the value $0.1190$ was taken).} 

\section{Modified beta functions and results}
\label{sec:fact}

In order to achieve the correct value $r_{\tau}=0.203$, and at the same time
preserve the holomorphic behavior of $F(z) \equiv a(Q^2)$, we follow in
principle the same line of reasoning as in Sec.~3 of Ref.~\cite{anpQCD1} and
Sec.~IV of Ref.~\cite{anpQCD2}.  
The idea is to replace in the beta function (\ref{ans1})
 \ba
f(Y) \mapsto  f(Y) f_{\rm fact}(Y) \ .
\label{modf}
\ea
Here, $f_{\rm fact}(Y)$ should be close to unity for the relevant values 
$Y \equiv a/a_0 \equiv F(z)/a_0$, i.e., for the values $Y$ around the interval
$(0,1)$ in the complex $Y$-plane, in order to obtain similar
results for $F(z)$ as in the case 
without an additional factor $f_{\rm fact}(Y)$.
This way there is a high probability that the coupling $F(z)$ for
the replaced (modified) beta function remains holomorphic (in the 
entire open stripe $- \pi < {\rm Im}z < + \pi$). We note that now the
conditions (\ref{con}) are applied to $f_{\rm new}(Y) \equiv f(Y) f_{\rm fact}(Y)$.

One of the consequences of this condition
is that the LB part of $r_{\tau}$ [cf.~Eq.~(\ref{LBrt2})] does not change much
by this replacement, and neither do the contour integrals  $I(\ta_{n+1})$
of the ${\rm N}^n{\rm LB}$ contribution 
[cf.~Eqs.~(\ref{rtman2})-(\ref{TnIta}) and (\ref{tan})].
The coefficient $T_1=1/12$ of the
NLB contribution remains scheme independent (and small), and therefore
also the NLB contribution does not change much (and remains small) 
under the mentioned modification.
In Table \ref{tabrtau1} we can see that for the original
schemes ($c_2^{(0)}, c_3^{(0)}, \ldots$), as defined by the
beta functions Eqs.~(\ref{ans1}) and (\ref{fP20})-(\ref{EE}) 
[cf.~also Eq.~(\ref{betaexp})], the 
${\rm N}^2{\rm LB}$ and ${\rm N}^3{\rm LB}$ contributions are too small
for achieving the correct value $r_{\tau}=0.203$. 
However, the coefficient $T_2$ of the ${\rm N}^2{\rm LB}$ contribution
depends strongly on the leading scheme parameter $c_2$, 
it changes linearly with $c_2$
scheme coefficient of the beta function [cf.~Eq.~(\ref{betaexp})]:\footnote{
This $c_2$-dependence of $T_2$ can be inferred from relations given in 
Appendices \ref{app1} and \ref{app2}: Eqs.~(\ref{tdn}) give us relations
between the $d_n$ and ${\widetilde d}_n$ coefficients of the expansions
(\ref{dexp}) of the (spacelike) Adler function $d(Q^2)$; Eq.~(\ref{Tn}) relates
these coefficients with the coefficients $T_n$ of the LB+bLB expansion
(\ref{rtman2}) of $r_{\tau}$, and Eq.~(\ref{FvsbFb}) gives us the
$c_2$-dependence of $d_2$ (and thus of $T_2$), where in Appendix \ref{app2} the Adler
function $d(Q^2)$ is a special case of function ${\cal F}(Q^2)$ 
with $\nu_0=1$ (and ${\cal F}_n = d_n$).}
$T_2(c_2) = T_2(c_2^{(0)}) - (c_2 - {\overline c}_2^{(0)})$. The idea
is then to  introduce in the beta function such $f_{\rm fact}(Y)$
which fulfills simultaneously the following two conditions:
\begin{itemize} 
\item [(a)] $f_{\rm fact}(Y) \approx 1$ 
in the sector of the $Y$ complex plane around the $(0,1)$ interval;
\item [(b)]
it decreases the value of $c_2$ to significantly lower values 
(from $c_2^{(0)} \sim - 10^1$ of Table \ref{tabrtau1} to $c_2 \sim - 10^2$).
\end{itemize}
The latter condition increases the $T_2$ coefficient and the
${\rm N}^2{\rm LO}$ by about one order of magnitude and thus allows us to
obtain the correct value $r_{\tau}=0.203$. 

In Refs.~\cite{anpQCD1,anpQCD2}, the $f_{\rm fact}(Y)$ functions which fulfilled the
two mentioned conditions were chosen essentially in the following form:
\be
f_{\rm fact}(Y) = 1 - \frac{K}{B} \left( \frac{ B Y^2}{1 + B Y^2} \right) \ ,
\label{ff1}
\ee
where $K \sim 10^1$ was needed to obtain $c_2 \sim - 10^2$ and thus the
correct $r_{\tau}=0.203$, and $B \gg K$ ($B \sim 10^3$) was needed to
keep $f_{\rm fact}(Y) \approx 1$ in the $Y$ complex plane around the
$(0,1)$ interval. In this way, the sum of the first four terms,
i.e., LB and  ${\rm N}^n{\rm LB}$ ($n=1,2,3$) contributions 
whose coefficients $T_n$ are exactly known, gave the correct value
 $r_{\tau}=0.203$. However, the next term (${\rm N}^4{\rm LB}$) was then
uncontrollably large: 
$|r_{\tau}({\rm N}^4{\rm LB})| \propto T_4 \sim c_4 \sim B K \sim 
10^6$-$10^7$. This took place due to the fact that the expansion of
$f_{\rm fact}(Y)$ in powers of $Y^2$ has a huge coefficient $B K$ at $Y^4$
($(a/a_0)^4$)
\be
f_{\rm fact}(Y)_{\rm exp} = 1 - K Y^2 + (B K) Y^4 - (B^2 K) Y^6 + \cdots
\label{ff1exp}
\ee

Within the present work we try to avoid this unwanted behavior in 
the following way: we modify
the expression (\ref{ff1}) for $f_{\rm fact}$ into an expression
${\cal F}_{\rm fact}$ such that the offending 
coefficients in the expansion (\ref{ff1exp}) disappear
\bes
\label{ff1ffexp}
\ba
f_{\rm fact}(Y) & \mapsto & {\cal F}_{\rm fact}(Y) \ , 
\label{ff1ffexpa}
\\
{\rm such \; that} \;\;\;\;
f_{\rm fact}(Y)_{\rm exp} & \mapsto & {\cal F}_{\rm fact}(Y)_{\rm exp} =
1 - K Y^2 + {\cal O}\left( (Y^2)^{P+1} \right) \ ,
\label{ff1ffexpb}
\ea
\ees
where the subscript ``exp'' denotes the expanded form of 
the corresponding function, $P$ is a large chosen integer, 
$K \sim 10^1$ as required by the mentioned condition (b),
and at the same time requiring that the condition  (a) survives:
${\cal F}_{\rm fact}(Y) \approx 1$
in the sector of the $Y$ complex plane around the $(0,1)$ interval,
in order to preserve analyticity of $F(z)$.
It turns out that such a transformation is possible 
[Eqs.~(\ref{cf})-(\ref{cf3})], as we show in the following.  
Let $g(\omega)$ be a function whose expansion around $\omega=0$ is

\be
g(\omega)_{\rm exp} = {\cal C}_1 \omega + {\cal C}_2 \omega^2 + \ldots \ .
\label{gomexp}
\ee
Consider the finite group of rotations of the complex plane given by
$$\omega\mapsto\exp\left(i\dfrac{2\pi}{P}k\right)\omega\quad\text{for }k=0,...,P-1.$$ Now consider the average of the images of $h(\omega) \equiv g(\omega)/\omega$ under this group of rotations of order $P$. We denote this average by $\widetilde h_P(\omega) \equiv \widetilde g_P(\omega)/\omega$:
\be
\frac{{\widetilde g}_P(\omega)}{\omega} \equiv {\widetilde h}_P(\omega)=\frac{1}{P}\sum_{k=0}^{P-1} h( \exp\left( i\dfrac{2\pi}{P}k\right) \omega ) \ .
 \label{tgom}
\ee
It is straightforward to verify that all the terms with
exponents that are not divisible by $P$ are annihilated.\footnote{ 
Using the expansion (\ref{gomexp}) in 
Eq.~(\ref{tgom}) leads to 
${\widetilde h}_P(\omega)_{\rm exp} = (1/P) \sum_{k=0}^{P-1} \sum_{N=0}^{\infty}
 C_{N+1} \exp(i 2 \pi k N/P) \omega^N =
\sum_{N=0}^{\infty} C_{N+1} \omega^N
\times (1/P) \sum_{k=0}^{P-1} \exp(i 2 \pi k N/P)$. 
When $N \not=0, P, 2P, \ldots$, we have
$\sum_{k=0}^{P-1} \exp(i 2 \pi k N/P)  = 
\left( \exp(i 2 \pi N) - 1 \right)/\left( \exp(i 2 \pi N/P) - 1 \right) = 0$;
and when $N=0, P, 2P, \ldots$, we have $\sum_{k=0}^{P-1} \exp(i 2 \pi k N/P) = P$.
This then leads to the expression (\ref{tgomexp}).}
We are left with
\be
{\widetilde g}_P(\omega)_{\rm exp} = {\cal C}_1 \omega + 
{\cal C}_{P+1} \omega^{P+1} + {\cal C}_{2 P +1} \omega^{2 P + 1} \ldots \ ,
\label{tgomexp}
\ee
i.e.,  the expansion series of ${\widetilde g}_P(\omega)$ has vanishing 
lowest-order terms (with the exception of the linear one).
When applying this approach to $g(\omega) \equiv f_{\rm fact}(Y) - 1$ 
of Eq.~(\ref{ff1}),
with $\omega=Y^2$, we obtain
\ba
{\cal F}_{\rm fact}(Y) & = &
1 - \frac{K}{B} \left( \frac{ B Y^2}{1 - (-B Y^2)^P} \right) \ ,
\label{cf0}
\ea
It turns out that the condition (a) is fulfilled only when $P$ is odd:
$P=2 N+1$ (because for $P=2 N$ beta function has a pole at 
small positive $Y=1/\sqrt{B}$). Consequently, we use $P=2N+1$
\bes
\label{cf}
\ba
{\cal F}_{\rm fact}(Y;N) & = &
1 - \frac{K}{B} \left( \frac{ B Y^2}{1 + (B Y^2)^{2 N+1}} \right) 
\label{cf1}
\\
& = & 
\frac{1 + (B Y^2)^{2 N+1} - K Y^2}{1 + (B Y^2)^{2 N+1}} \ ,
\label{cf2}
\ea
\ees
and the expansion around $Y=0$ is
\be
\Rightarrow \; {\cal F}_{\rm fact}(Y;N)_{\rm exp} =  
1 - K Y^2 + \frac{K}{B} \left[ (B Y^2)^{2 N+2} -
(B Y^2)^{4 N+3} + (B Y^2)^{6 N+4} \cdots \right] \ .
\label{cf3}
\ee
We recall that $K \sim 10^1$ and $B \gg K$. The new considered beta 
functions are now, according to Eqs.~(\ref{ans1})-(\ref{con}), (\ref{modf}),
(\ref{ff1ffexp})
\be
\beta(F) = -\beta_0 F^2 (1 - Y) f(Y) {\cal F}_{\rm fact}(Y; N) \ ,
\label{betanew}
\ee
where $F \equiv a(Q^2)$, $Y \equiv F/a_0 \equiv a/a_0$. For $f(Y)$ 
we stick to the original options 
Eqs.~(\ref{fP20}) ($r_2=0$) or (\ref{fP11}) or (\ref{EE}).
We note that the conditions (\ref{con}) are applied now to
 $f_{\rm new}(Y) \equiv f(Y) {\cal F}_{\rm fact}(Y)$.

Here we recall once more that the physical condition of obtaining large 
enough value of $r_{\tau}$ ($\approx 0.203$) imposed on us: (1) the condition 
of having a large value $K \sim 10^1$ in the expansion 
${\cal F}_{\rm fact}(Y;N)_{\rm exp} = 1 - K Y^2 + {\cal O}(Y^3)$
with the coefficients at higher powers of $Y$ under control;
(2) and simultaneously the condition ${\cal F}_{\rm fact}(Y;N) \approx 1$ in the
sector of the $Y$ complex plane around the $(0,1)$ interval. 
Mathematically, these two conditions tend to be in general
in conflict, which explains why it was so difficult to obtain
a solution, such as Eqs.~(\ref{cf})-(\ref{cf3}), 
reconciling both of them.\footnote{The problem of too low
value of $r_{\tau}$ was already encountered and criticized
in Ref.~\cite{Geshkenbein} in the case of the analytic (and nonperturbative)
QCD model APT of Refs.~~\cite{ShS1,ShS2,MS,Sh1Sh2a,Sh1Sh2b}. 
The formal reason for
the problem of (the tendency to) too low $r_{\tau}$ 
was identified in Ref.~\cite{Geshkenbein} in the fact that in analytic models
the Landau cut of the coupling at positive $Q^2 \equiv - \sigma$
($0 < Q^2 \leq \Lambda_{\rm L.}^2$) 
is missing and therefore the integral for $r_{\tau}$ with positive integrand,
e.g. Eq.~(\ref{LBrt2}),
has a ``missing'' integration interval $-\Lambda_{\rm L.}^2 \leq \sigma < 0$
[$\sigma \equiv t e^{\cal {\overline C}} m_{\tau}^2$ in Eq.~(\ref{LBrt2})]
along the Landau cut and thus gives in general too small $r_{\tau}$.}

In Refs.~\cite{anpQCD1,anpQCD2} only the modification with 
${\cal F}_{\rm fact}(Y;N=0)$ was investigated. 
We now perform the numerical integration
of the RGE (\ref{RGExy}) for many different $N$, as high as $N=25$,
and adjust $B$ ($\sim 10^3$) and $K$ ($\sim 10^1$) so that
the correct value of $r_{\tau}$ is obtained by adding the first four
terms. The higher terms $r_{\tau}({\rm N}^{n}{\rm LB})$ ($n \geq 4$)
are now under control, they are estimated to contribute less than 0.001.
And the holomorphic behavior of $F(z) \equiv a(Q^2)$ is preserved when
$N$ increases. In Table \ref{tabrtau2} we present the results
analogous to those in Table \ref{tabrtau1}, but now with
$f(Y) \mapsto f(Y) {\cal F}_{\rm fact}(Y;N)$ 
for the choices $N=0$ and $N=25$, again for the three cases of 
$f(Y)$: Eq.~(\ref{fP20}) with $r_2=0$ (P[1/0]); Eq.~(\ref{fP11})
with $u_1=-0.1$ (P[1/1]);  Eq.~(\ref{EE}) with $Y_1=0.1$, $k_1=10.0$,
$k_2=11.0$. 
\begin{table}
\caption{Same as in Table \ref{tabrtau1}, but now
beta function is changed by the substitution
$f(Y) \mapsto f(Y) {\cal F}_{\rm fact}(Y;N)$,
Eqs.~(\ref{cf})-(\ref{betanew}), where $N=0$ and $N=25$.
Included are new beta function coefficients $c_j$ ($j=2,3,4$),
$a_{\rm in}$ and $a_0$ values, and the threshold mass value $M_{\rm thr}$ (in GeV).}
\label{tabrtau2} 
\begin{ruledtabular}
\begin{tabular}{llll|lllllccclll}
$f$ & $N$ & $B$ & $K$ & $r_{\tau}:$ LB  & NLB & ${\rm N}^2{\rm LB}$ & ${\rm N}^3{\rm LB}$ & sum & 
$c_2$ & $c_3$ & $c_4$  & $a_{\rm in}$ & $a_0$ &  $M_{\rm thr}$ [GeV]
\\
\hline
P[1/0] & 0 & 5000 & 6.8 & 0.1055 & 0.0006 & 0.0911 & 0.0058 & 0.2029 & -224.67 & -333.73 & $2.59 \times 10^7$ & 0.0574 & 0.1903 & 0.159
\\
P[1/0] & 25 & 2100 & 6.8 & 0.1054 & 0.0006 & 0.0913 & 0.0058 & 0.2030 & -225.19 & -334.52 & 6966.7 & 0.0574 & 0.1901 & 0.159 
\\
\hline
P[1/1] & 0 & 4650 & 7.3 & 0.1056 & 0.0006 & 0.0910 & 0.0058 & 0.2030 & -220.89 & -307.36 & $2.14 \times 10^7$ & 0.0573 & 0.1994 & 0.149 
\\ 
P[1/1] & 25 & 2000 & 7.3 & 0.1055 & 0.0006 & 0.0914 & 0.0059 & 0.2032 & -221.49 &  -308.18 & 6895.3 & 0.0573 & 0.1992 & 0.149
\\
\hline
EE & 0 & 1150 & 5.6 & 0.1235 & 0.0007 & 0.0690 & 0.0098 & 0.2029 & -110.45 & -333.35 & $2.04 \times 10^6$ & 0.0608 & 0.2369 & 0.196
\\
EE  & 25 & 500 & 5.6 & 0.1231 & 0.0007 & 0.0694 & 0.0098 & 0.2029 & -111.32 & -336.33 & 441.47 & 0.0606 & 0.2360 & 0.196
\end{tabular}
\end{ruledtabular}
\end{table}
Note how strongly $c_4$ gets suppressed when going from $N=0$ to $N=25$.
There is some freedom of varying $K$ and $B$ so that
$r_{\tau}=0.203$ is obtained. However, when we decrease $B$ (at a given $N$),
$K$ has to be increased somewhat, and the convergence properties of the
first four terms in the sum for $r_{\tau}$ deteriorate: LB contribution
decreases, and the ${\rm N}^2{\rm LB}$ contribution increases. Since we want
to have the first (LB) term clearly dominant, we are forced to 
use relatively high values of $B$ ($\sim 10^3$). Though, when $N$ increases,
we can use somewhat lower values of $B$ while still maintaining the
same convergence quality.

It turns out that Figures \ref{figP20bt}(a), \ref{figEEbt}(a), \ref{figbet}
and \ref{figrho} change only a little when the modifications with the
parameters ($N, B, K$) 
as given in Table \ref{tabrtau2} are peformed. Most importantly,
the holomorphic behavior of $F(z) \equiv a(Q^2)$, as signalled by 
Figs.~\ref{figP20bt}(a) and \ref{figEEbt}(a) in the nonmodified case ($K=0$),
is preserved.\footnote{
This is true also in the other considered case:
$f(Y) = {\rm P}[1/1](Y)$ with $u_1=-1$.} 
\begin{figure}[htb] 
\begin{minipage}[b]{.49\linewidth}
\centering\includegraphics[width=85mm]{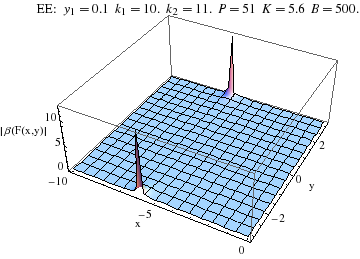}
\end{minipage}
\begin{minipage}[b]{.49\linewidth}
\centering\includegraphics[width=85mm]{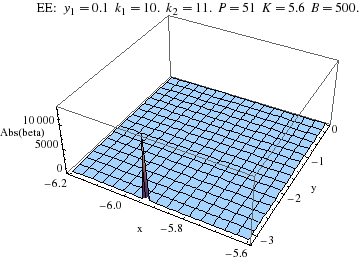}
\end{minipage}
 \caption{\footnotesize  $|\beta(F(z))|$ as a function of
$z \equiv \ln(Q^2/Q^2_{\rm in}) =x+i y$, as in Fig.~\ref{figEEbt}(a), but now with the modified
beta function with $P=51$ ($=2 N +1$), $K=5.6$ and $B=500$; the
right-hand figure has more details around the pole. No Landau singularities 
(i.e., singularities inside the $z$ stripe) appear.}
\label{figEEP51bt}
 \end{figure}
\begin{figure}[htb] 
\begin{minipage}[b]{.49\linewidth}
\centering\includegraphics[width=85mm]{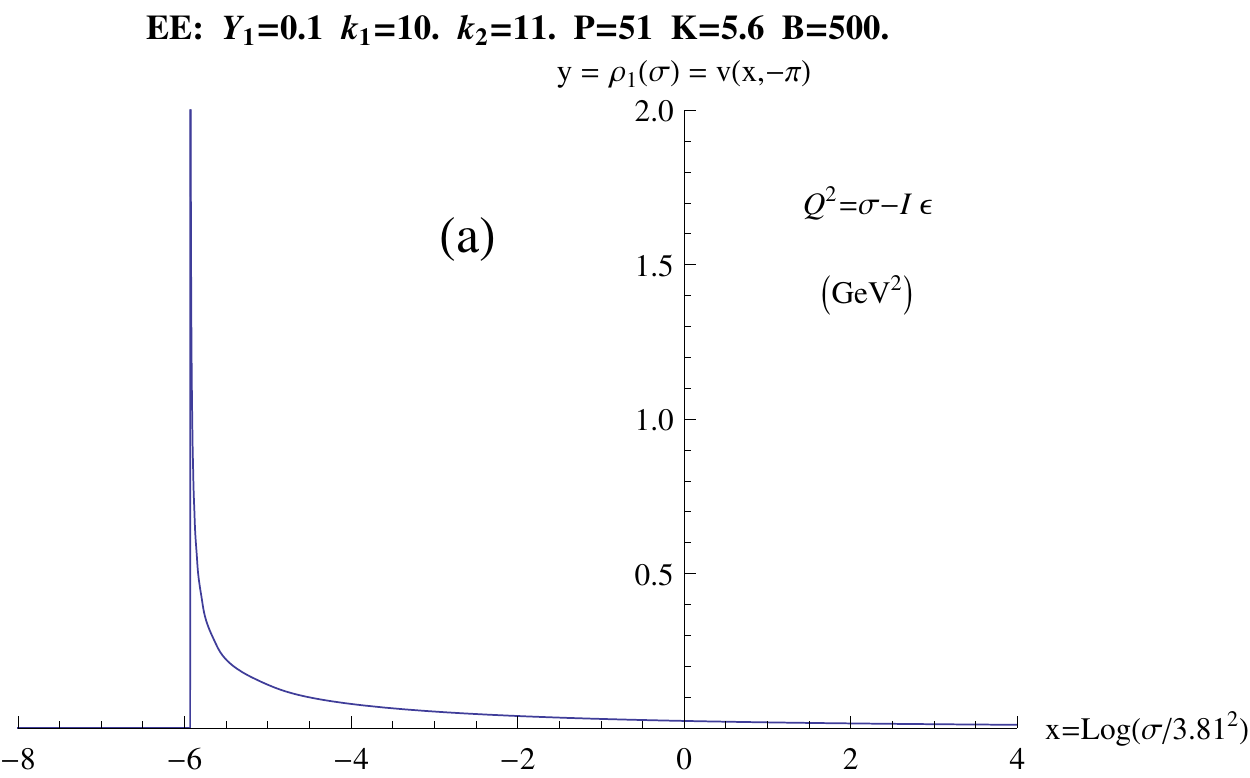}
\end{minipage}
\begin{minipage}[b]{.49\linewidth}
\centering\includegraphics[width=85mm]{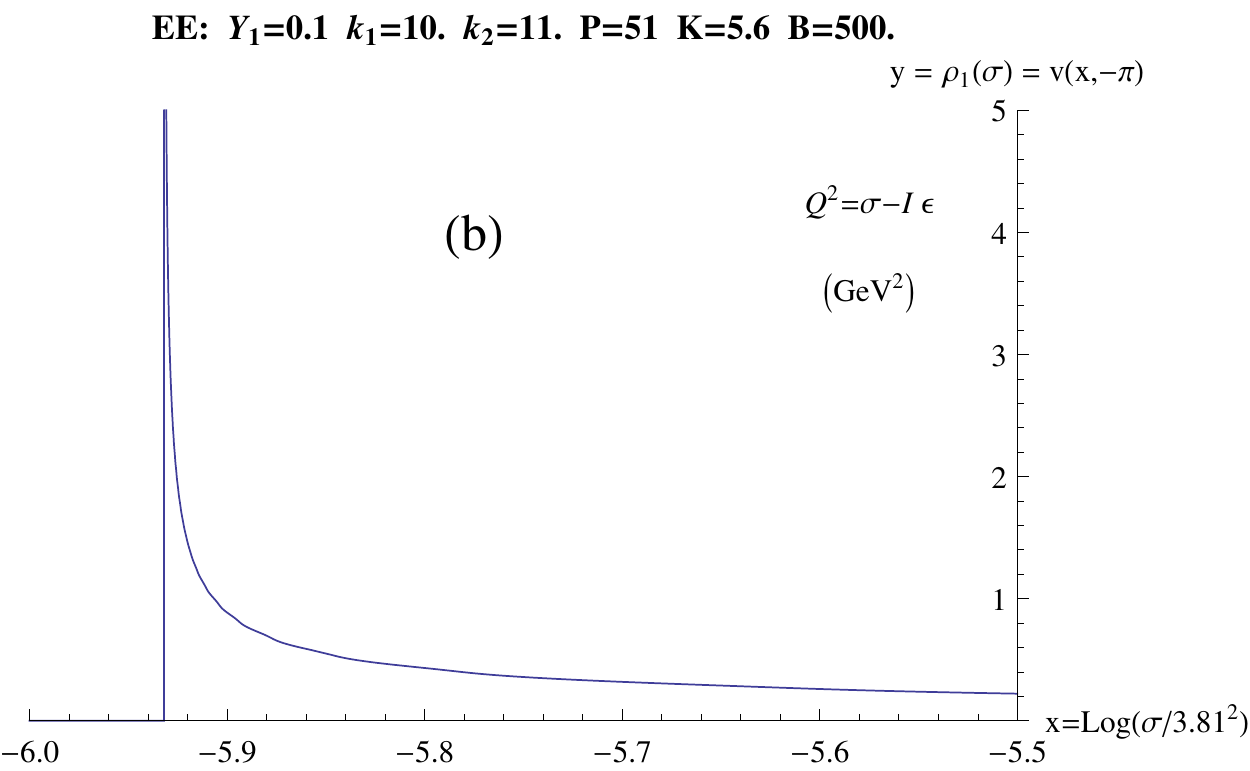}
\end{minipage}
 \caption{\footnotesize  (a) The discontinuity function
$\rho_1(\sigma) = {\rm Im} \ a(Q^2=-\sigma - i \epsilon) =
{\rm Im} F(z=x - i \pi) = v(x,y=-\pi)$ as a function of
$x = {\rm Re}(z) = \ln(\sigma/Q^2_{\rm in})$, for the EE case with
modified beta function ($P=51$, i.e., $N=25$); (b) enlarged
picture around the threshold.}
\label{figrhoEEP51}
 \end{figure}
\begin{figure}[htb] 
\begin{minipage}[b]{.49\linewidth}
\centering\includegraphics[width=85mm]{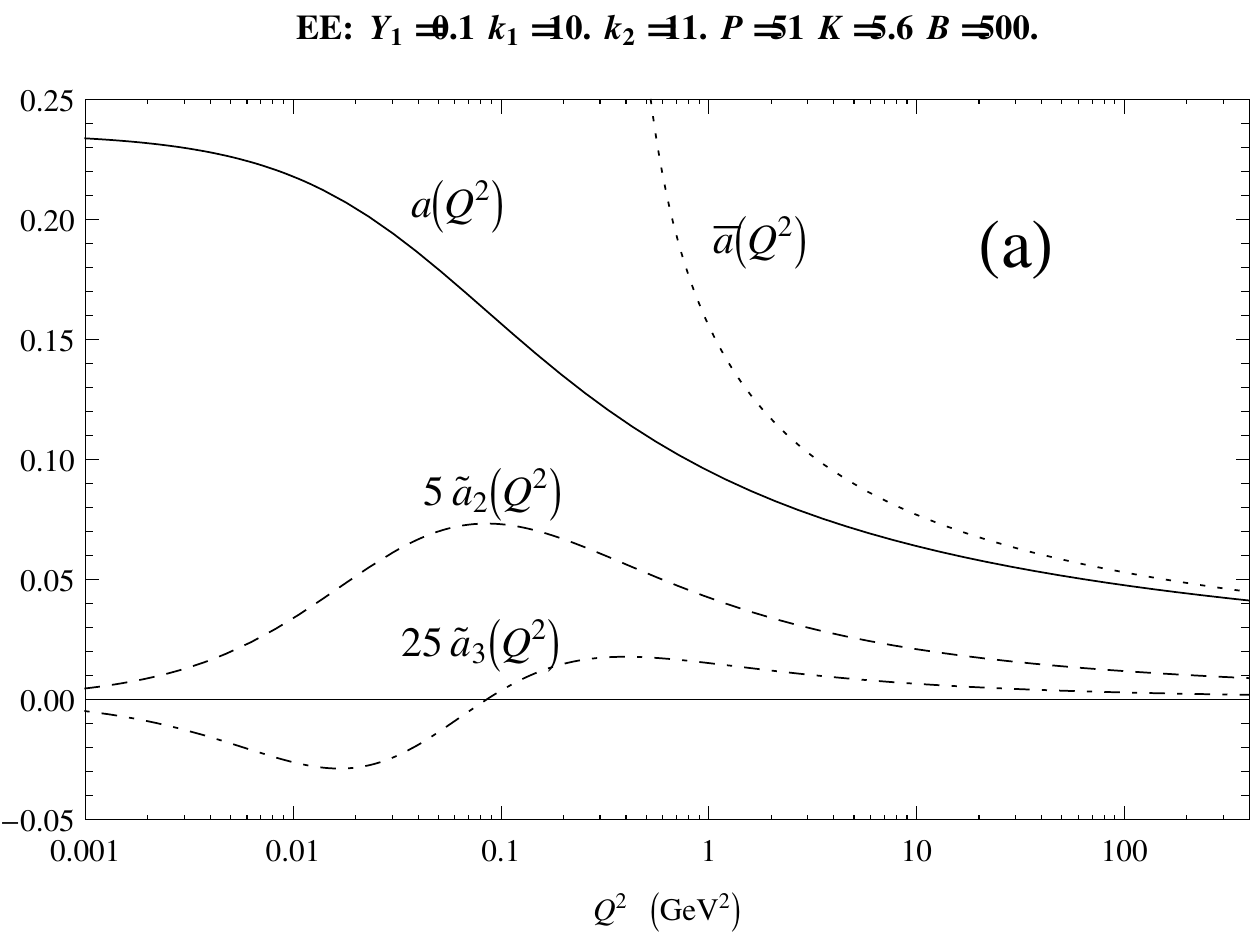}
\end{minipage}
\begin{minipage}[b]{.49\linewidth}
\centering\includegraphics[width=85mm]{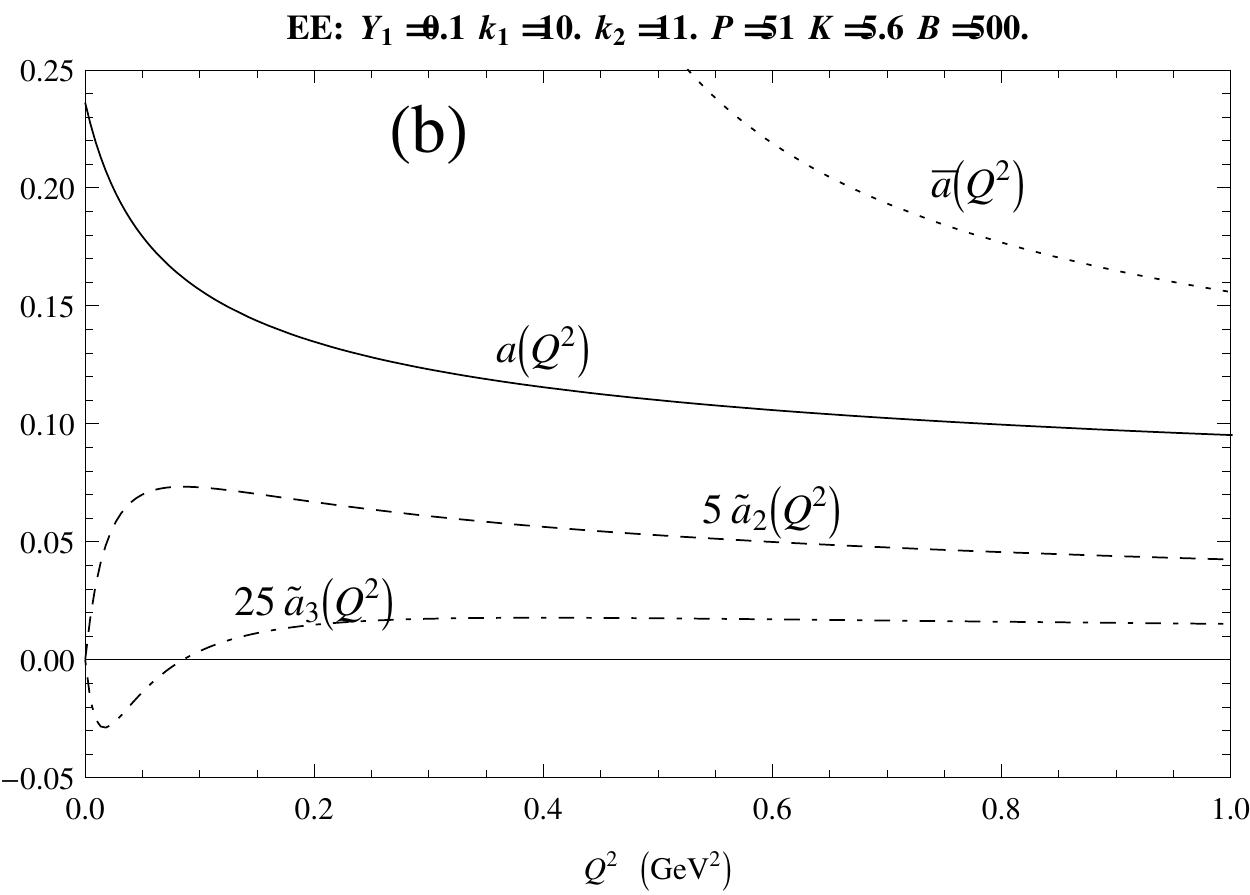}
\end{minipage}
\vspace{-0.2cm}
 \caption{\footnotesize  (a) The running coupling
$a(Q^2)$ as a function of positive $Q^2$, for the EE case with
modified beta function ($P=51$, i.e., $N=25$), when $Q^2$ is on a
logarithmic scale; included are the logarithmic
derivatives $\ta_2(Q^2)$ and $\ta_3(Q^2)$, Eq.~(\ref{tan}), rescaled by factors
$5$ and $25$ for better visibility; for comparison, the corresponding
$\MSbar$ coupling ${\overline a}(Q^2)$ is included; (b) same as in Fig.~(a),
but on linear $Q^2$ scale for low positive $Q^2$.}
\label{figaEEP51}
 \end{figure}
For this, we present in Figs.~\ref{figEEP51bt} the behavior of the 
function $|\beta(F(z))|$ for the EE case with $N=25$ 
(i.e., $P \equiv 2 N+1 = 51$),
which clearly indicates that there are no Landau singularities. 
In addition, in Figs.~\ref{figrhoEEP51} we present for this
case the discontinuity function $\rho_1(\sigma) \equiv {\rm Im} a(-\sigma-i \epsilon)$ as a function of 
$\ln(\sigma/Q^2_{\rm in})$, and in Figs.~\ref{figaEEP51} the running
coupling $a(Q^2)$, and its logarithmic derivatives $\ta_2(Q^2)$ and $\ta_3(Q^2)$
[defined in Eq.~(\ref{tan})], for positive $Q^2$.

While the considered perturbative beta functions modified 
by the factor (\ref{cf}) give us simultaneously the correct value of
$r_{\tau}$ and perturbative holomorphic coupling $a(Q^2)$, one may
worry that the introduction of the large coefficients 
$\sim B^{2 N+1}$ (where $B \sim 10^3$)
in the beta function [cf.~Eq.~(\ref{cf3})] represents an anomalous 
mass independent renormalization scheme, in the sense that the
growth of the coefficients of beta function at large order becomes
responsible for a growth of the coefficients of the physical
spacelike physical quantities which is faster than the growth coming from
the leading (UV or IR) renormalon. Here we will argue that if
$P \equiv 2 N+1$ in (\ref{cf}) is large enough, e.g. $P \geq 51$, then the
renormalon growth of the coefficients will dominate over the
growth from the beta function coefficients. For example,
in the case of the timelike quantity $r_{\tau}$, the underlying
spacelike quantity is Adler function $d(Q^2)$
(with $N_f=3$), cf.~Eq.~(\ref{rtaucont}).

Let us consider a general spacelike physical quantity ${\cal D}(Q^2)$,
whose expansion is
\be
{\cal D}(Q^2) = a(Q^2) + d_1 a(Q^2)^2 + \cdots + d_n a(Q^2)^{n+1} + \cdots 
\label{Dexp}
\ee
The expansion of its Borel transform is
\be
B_{\cal D}(b) = 1 + \frac{d_1}{1! \beta_0} b + \cdots \frac{d_n}{n! \beta_0^n} b^n + \cdots \ .
\label{BD}
\ee
It turns out that this function can have 
(renormalon) poles only at nonzero integer values
$b = \pm 1, \pm 2, \ldots$, cf.~Ref.~\cite{Beneke:1998ui}. 
The closer the renormalon pole is to the origin, 
the faster is the increase of the coefficients $d_n$
with $n$. Let us assume that the pole is at either $b=1$ or $b=-1$.
Then, in the large-$\beta_0$ approximation, the coefficients $d_n$
behave at large $n$ as
\be
|d_n| \approx {\cal K} n! \beta_0^n \sim n! \beta_0^n \ ,
\label{dnren}
\ee
where ${\cal K} \sim 1$.\footnote{
We recall that the renormalon problem is reflected in this
growth of the coefficients $d_n$, and is not related with the
existence or nonexistence of the Landau singularities of the running coupling
$a(Q^2)$. The question whether the Landau singularities appear or not is a
problem of the running coupling and its beta function only, 
cf.~comments at the end of Sec.~2.2 of the review Ref.~\cite{Beneke:1998ui}.}
We recall that in our notation, $a \equiv \alpha_s/\pi$ and $\beta_0 = (1/4)(11 - 2 N_f/3)$, i.e., $\beta_0=9/4$ when $N_f=3$. 
On the other hand, the perturbative
scheme independence of the physical quantity ${\cal D}(Q^2)$ 
implies that the coefficient $d_n$ has a specific dependence on
the scheme coefficients $c_2, c_3, \ldots, c_n$ of the beta function expansion 
(\ref{betaexp}). In particular, the dependence on $c_n$ is
\be
d_n(c_n, c_{n-1},\ldots,c_2) = - \frac{1}{n-1} c_n + f_n(c_{n-1},\ldots,c_2) \ .
\label{dncn}
\ee
On the other hand, it is straightforward to check that for the
beta function (\ref{betanew}) modified by the factor (\ref{cf}), with $B \sim 10^3$,
$K \sim 10^1$ and $P=2 N+1$, the first anomalous (large) beta coefficient is
\be
c_{2 P + 2} \approx (-1)^{P+1} \frac{K}{a_0^2} \left( \frac{B}{a_0^2} \right)^P \ .
\label{c2p2}
\ee
This implies, together with Eq.~(\ref{dncn}), that the first
$d_n$ with anomalously large beta coefficient is
\be
d_{2 P + 2} = \frac{(-1)^P}{2 P+1} \frac{K}{a_0^2} \left( \frac{B}{a_0^2} \right)^P
+ \ldots \ ,
\label{dnc}
\ee
where the dots stand for a contribution that is independent of the 
anomalous scheme parameter $B$. We require that the contribution 
(\ref{dnc}) of the anomalous scheme to the coefficient 
$d_{2P + 2}$ is less than the 
contribution of the $b = \pm 1$ (leading) renormalon (\ref{dnren}),
and this implies
\ba
\frac{1}{2 P+1} \frac{K}{a_0^2} \left( \frac{B}{a_0^2} \right)^P &<&
(2 P +2)! \beta_0^{2 P+2}  
\nonumber\\
\Rightarrow \; rat(P) & \equiv & \frac{K B^p}{ a_0^{2P+2} (2 P+1) (2P+2)!
\beta_0^{2 P+2}} < 1
\label{ratcon}
\ea
In the considered case of EE with $N=25$ ($P=51$), i.e., the 
last line of Table \ref{tabrtau2}, it turns out that this ratio
$rat(P)$ is $0.09$, i.e., the $b=\pm 1$ renormalon growth of the coefficients
$d_n$ clearly dominates over the growth of $d_n$ coming from the scheme.
For the other two cases (P[1/1] and P[1/0]) in Table \ref{tabrtau2},
this ratio is still huge, primarily due to the larger value of
$B$ ($\approx 2000$), and a significantly larger value of $P$
is needed.\footnote{
If we use in the numerical integration of the RGE a very large value of $P$
($P > 100$), the calculation (with MATHEMATICA) becomes either very
time-consuming, or it does not perform due to overflow problems.}
If the renormalon effects are accounted for beyond the large-$\beta_0$
approximation, the growth of the coefficients $d_n$ becomes even
slightly faster \cite{Beneke:1998ui}. Furthermore, in the specific
case of $r_{\tau}$, where the underlying coefficients $d_n$ are those of
Adler function, the leading renormalon is at $b=-1$ and it is double
(quadratic), so that the $|d_n|$ coefficients grow even slightly faster
than in Eq.~(\ref{dnren}), as $\sim (n+1)! \beta_0^n$.

\section{Borel sum rules in V+A channel of tau lepton semihadronic decays}
\label{sec:BSR}

In this Section we extract the four and six-dimensional condensates 
$\langle O_4^{\rm (V+A)} \rangle = (1/6) \langle a G^{\alpha}_{\mu \nu} G^{\alpha}_{\mu \nu} \rangle$ and $\langle  O_6^{\rm (V+A)} \rangle$ 
appearing in the Operator product expansion (OPE) of the $V+A$
quark current correlator $\Pi(Q^2)$, 
based on the measurements of the $\tau$-lepton
semihadronic decays. We use the Borel transform sum rules, following 
closely the approach of Ref.~\cite{anOPE}, where the evaluation was 
performed for the (nonperturbative) $2$-delta analytic QCD 
(2$\delta$anQCD) model 
of Ref.~\cite{2danQCD}. That approach followed the Borel transform
sum rule methods of Refs.~\cite{Geshkenbein,Ioffe}. 
We outline only the main features of the
approach and refer for details of the approach to Ref.~\cite{anOPE}
and Ref.~\cite{Ioffe}.

The starting point is the identity (sum rule)
\be
\int_0^{\sigma_0} d \sigma g(-\sigma) \omega_{\rm exp}(\sigma)  =
-i \pi  \oint_{|Q^2|=\sigma_0} d Q^2 g(Q^2) \Pi_{\rm th}(Q^2)  \ ,
\label{sr1}
\ee
where $g(Q^2)$ is an analytic (holomorphic) function in the 
entire $Q^2$ complex plane, which characterizes the specific sum rule.
The contour integration on the right-hand side 
is in the counterclockwise direction, and
$\omega(\sigma)$ is the spectral function of the
$V+A$ quark current correlator function $\Pi(Q^2)$
\be
\omega(\sigma) \equiv 2 \pi \; {\rm Im} \ \Pi(Q^2=-\sigma - i \epsilon) \ .
\label{om1}
\ee
The identity (\ref{sr1}) comes from applying the Cauchy theorem to 
the function $g(Q^2) \Pi(Q^2)$ and accounting for the correct holomorphic
behavior of the correlator $\Pi(Q^2)$ as required by the general principles
of quantum field theories. The same type of holomorphic behavior is
respected by the QCD running couplings $a(Q^2)$ and $\ta_{n}(Q^2)$
in the schemes considered in this work. Therefore, the theoretically
evaluated correlators $\Pi_{\rm th}(Q^2)$ 
[$\Leftrightarrow {\cal D}_{\rm Adl}(Q^2)$],
at each order of truncation in the considered holomorphic schemes,
have the analytic behavior consistent with the identity (\ref{sr1}).

In the present case, we are interested in $V+A$ channel of $\tau$ lepton 
semihadronic nonstrange decays. The experimental spectral function 
$\omega_{\rm exp}(\sigma)$ on the left-hand side of 
the sum rule (\ref{sr1}) is obtained from the invariant-mass spectra of the
$\tau$ lepton strangeless decays with the squared invariant mass $\sigma$
in the interval $0 < \sigma < \sigma_0$.  Our analysis here is based on the
data of ALEPH Collaboration  \cite{ALEPH1,ALEPH2a,ALEPH2b}. 
On the right-hand side
of the sum rule is the correlator function
$\Pi(Q^2)$, which is theoretically evaluated with OPE
\be
\Pi(Q^2) =  - \frac{1}{2 \pi^2} \ln(Q^2/\mu^2) + 
\Pi(Q^2;D\!=\!0)
+ \sum_{n \geq 2} \frac{ \langle O_{2n} \rangle}{(Q^2)^n} \left(
1 + {\cal C}_n a(Q^2) \right) \ .
\label{OPE1}
\ee
Note that $n=D/2$ where $D$ denotes the operator dimension of the
local operators contributing to the OPE of $\Pi(Q^2)$. 
The $D=2$ ($n=1$) term is 
proportional to the current masses of $u$ and $d$ 
quarks, and is negligible. For us the relevant terms are $D=4,6$ ($n=2,3$).
Further, it can be checked that the terms proportional to  ${\cal C}_n a(Q^2)$ 
will be negligible in the Borel sum rules applied here (cf.~footnote 20 of 
Ref.~\cite{anOPE}).

For the evaluation of the right-hand side of the sum rule (\ref{sr1}),
it turns out convenient to integrate it by parts
\be
\int_0^{\sigma_0} d \sigma g(-\sigma) \omega_{\rm exp}(\sigma) =
- \frac{i}{2 \pi}  \int_{\phi=-\pi}^{\pi} \frac{d Q^2}{Q^2}
{\cal D}_{\rm Adl}(Q^2) \left[{\cal G}(Q^2) - {\cal G}(-\sigma_0) \right] {\big |}_{Q^2 = \sigma_0 \exp(i \phi)} \ ,
\label{sr2}
\ee
where ${\cal G}$ is any function satisfying
\be
\frac{d {\cal G}(Q^2)}{d Q^2} = g(Q^2) \ ,
\label{Ff}
\ee
and ${\cal D}_{\rm Adl}(Q^2)$ is the full massless Adler function
\bes
\label{dOPE}
\bea
{\cal D}_{\rm Adl}(Q^2) &\equiv&  - 2 \pi^2 \frac{d \Pi(Q^2)}{d \ln Q^2} 
\label{Adl1}
\\
& = & 1 + d(Q^2) + 2 \pi^2 \sum_{n \geq 2}
 \frac{ n \langle O_{2n} \rangle}{(Q^2)^n}  \ ,
\label{OPE2}
\eea
\ees
where the terms with ${\cal C}_n a(Q^2)$ were neglected, as mentioned earlier.
The dimension $D=0$ part of the correlator is directly related to the
(strangeless and massless) canonical Adler function $d(Q^2)$ of
Appendix \ref{app1} [Eqs.~(\ref{dexp}), (\ref{dLB})-(\ref{dbLB})] 
\be
d(Q^2) =  - 2 \pi^2 \frac{d \Pi(Q^2;D\!=\!0)}{d \ln Q^2} \ .
\label{Adl2}
\ee

In the sum rule (\ref{sr1}), the analytic function $g(Q^2)$ is usually
taken to be either an exponential function $\propto \exp(Q^2/M^2)$
(Borel sum rules, Refs.~\cite{Ioffe,Ioffe:2000ns}), or a Gaussian function
$\propto \exp( (Q^2/M^2)^2 )$ (Gaussian sum rules, Ref.~\cite{Ioffe:2000ns}),
or powers $\propto (Q^2)^N$. The integrals of the latter approach
are called moments, and the corresponding sum rules are usually called 
finite energy sum rules, 
cf.~Refs.~\cite{Ioffe:2000ns,vsr1,vsr2,vsr3,vsr4,vsr5,vsr6,vsr7,vsr8}.

Although the approach with moments is more widely used in the literature, 
we will calculate the Borel transforms, i.e., we will apply the Borel sum rules 
\cite{Ioffe}, the main reason being that we already have experience
and acquired confidence in such calculations, cf.~Ref.~\cite{anOPE}. 
Nonetheless, it would
be interesting to apply in the future the moment approach to the considered
holomorphic schemes.

Therefore, our choice for $g(Q^2)$ here will be
\be
g(Q^2) = \frac{1}{M^2} \exp( Q^2/M^2) \ , \qquad
{\cal G}(Q^2) = \exp( Q^2/M^2) \ ,
\label{fFBorel}
\ee
where $M^2$ are chosen complex scales with 
${\rm Re}(M^2)>0$.
The expressions in the sum
rules (\ref{sr1}) and (\ref{sr2}) become Borel transforms $B(M^2)$,
and we choose there for the upper integration bound 
the maximal possible value $\sigma_0 = m_{\tau}^2$
($\approx 3.16 \ {\rm GeV}^2$).\footnote{If $\sigma_0$ is taken
well below $m_{\tau}^2$, the duality-violating effects become
important and must be taken into account, see Refs.~\cite{DV1,DV2}.}
The Borel sum rule thus has the form
\be
B_{\rm exp}(M^2) =  B_{\rm th}(M^2) \ ,
\label{sr3a}
\ee
where
\bes
\label{sr3}
\bea
B_{\rm exp}(M^2) &\equiv& \int_0^{m_{\tau}^2} 
\frac{d \sigma}{M^2} \; \exp( - \sigma/M^2) \omega_{\rm exp}(\sigma) \ ,
\label{sr3b}
\\
B_{\rm th}(M^2) &\equiv& B(M^2;D\!=\!0) + 2 \pi^2 \sum_{n \geq 2}
 \frac{ \langle O_{2n} \rangle}{ (n-1)! \; (M^2)^n} \ ,
\label{sr3c}
\eea
\ees
The $D=0$ part is
\bea
\lefteqn{
B(M^2;D\!=\!0) =  \left( 1 - \exp(-m_{\tau}^2/M^2) \right) }
\nonumber\\
&&+
\frac{1}{2 \pi}\int_{-\pi}^{\pi}
d \phi \; d(Q^2\!=\!m_{\tau}^2 e^{i \phi}) \left[ 
\exp \left( \frac{m_{\tau}^2 e^{i \phi}}{M^2} \right) -
\exp \left( - \frac{m_{\tau}^2}{M^2} \right) \right] .
\label{BD0}
\eea
For small positive ${\rm Re}(M^2)$, the Borel transform suppresses strongly
the contributions of $\omega_{\rm exp}(\sigma)$ at high energies 
(high $\sigma$) where the
experimental errors are larger. Further, the OPE higher dimension terms
are suppressed in the Borel transform by a factor $1/(n-1)!$.
In the real part of the Borel transform, OPE term contributions
of specific dimension $D$ are eliminated if the complex scales $M^2$ are 
chosen along specific rays $M^2 = |M^2| \exp(i \psi)$ in the complex plane. 
This facilitates the determination of the remaining condensates
$\langle O_{2 n} \rangle$ by comparing 
the theoretical expressions with the experimental
values ${\rm Re} {\rm B}_{\rm exp}(|M^2| \exp(i \psi))$.
For example, if $\psi=\pi/6, \pi/4$, then $D=6, 4$ terms do
not contribute, respectively, because ${\rm Re}(\exp(i \pi/2)) = 0$.
Therefore, when ignoring terms with $D > 6$, we have
\bes
\label{pi64}
\bea
{\rm Re} B_{\rm exp}(|M|^2 e^{i \pi/6} ) &=& 
{\rm Re} B(|M|^2 e^{i \pi/6};D\!=\!0) + \pi^2 \frac{ \langle O_4 \rangle}{|M|^4}
\ ,
\label{pi6}
\\
{\rm Re} B_{\rm exp}(|M|^2 e^{i \pi/4} ) &=& 
{\rm Re} B(|M|^2 e^{i \pi/4};D\!=\!0) 
- \pi^2 \frac{ \langle O_6 \rangle}{\sqrt{2} |M|^6}
\ ,
\label{pi4}
\eea
\ees
We note that in the considered $V+A$ channel, the $D=4$ operator is
proportional to the gluon condensate
\be
\langle O_4^{(V+A)} \rangle = \frac{1}{6} \langle a G^{\alpha}_{\mu \nu}
G^{\alpha}_{\mu \nu} \rangle \ ,
\label{aGG1}
\ee
and the $D=6$ operator, in the vacuum saturation approximation,
is nonnegative and proportional to the square of the quark-antiquark condensate
\cite{Geshkenbein,Ioffe}
\be
\langle O_6^{(V+A)} \rangle   \approx  
\frac{128 \pi^2 }{81} a \langle {\overline q} q \rangle^2 \ .
\label{aqq1}
\ee
Here, as throughout this work, the notation $a \equiv \alpha_s/\pi$ is used.

The experimental values ${\rm Re} B_{\rm exp}(|M^2| \exp(i \psi))$ we use here
are those of Figs.~4 and 5(a),(b) of Ref.~\cite{Geshkenbein} (they were used also in Ref.~\cite{anOPE}), which are based on the values $\omega_{\rm exp}(\sigma)$
of the ALEPH 1998 data \cite{ALEPH1}. For the theoretical values 
${\rm Re} B_{\rm th}(|M^2| \exp(i \psi))$, 
the evaluation of the contour integrals (\ref{BD0}) of the canonical 
Adler function $d(Q^2)$ was performed with renormalization scale $\mu^2=Q^2$,
in the EE renormalization scheme: beta function
(\ref{betanew}) with $f(Y)$ of Eq.~(\ref{EE}) 
[with $Y_1=0.1$; $k_1=10.$; $k_2=11.$] and ${\cal F}_{\rm fact}$
of the form (\ref{cf}) with $B=500$, $K=5.6$ and 
$P \equiv 2 N+1 = 51$. For completeness, we write down here also values of
the other beta function parameters $Y_2$ and $a_0 \equiv \A_1(0)$ 
in this specific EE scheme, obtained numerically through the conditions 
(\ref{conc1})-(\ref{conanQ0}) applied to 
 $f_{\rm new}(Y) \equiv f(Y) {\cal F}_{\rm fact}(Y)$:
$Y_2=0.1839408532$ and $a_0=0.2360296246$.
We choose
this scheme because, as shown hitherto in this work, it represents
an analytic (holomorphic) perturbative QCD, gives the correct value
of $r_{\tau}$ decay ratio (cf.~the last line in Table \ref{tabrtau2})
and the growth of the coefficients $d_n$ with rising $n$ is dominated by the
leading renormalon $b=\pm 1$ (cf.~the previous Section)
and not by the beta function. 

Furthermore, $r_{\tau}$
was calculated in the LB+bLB approach, which is applicable if the
running coupling $a(Q^2)$ is holomorphic and which uses the maximal amount
of the presently available information on the perturbation coefficients 
of Adler function, and is thus considered as one of 
the most effective resummation
approaches for the $\tau$ decay physics quantities. Therefore,
we apply the LB+bLB approach also in the calculation of the contour
integrals (\ref{BD0}) for the Borel sum rules. We refer to Appendix
\ref{app3} for some formal details of the LB+bLB approach to the
Borel sum rules (analogous to Appendix \ref{app1} which explains
the calculation of $r_{\tau}$ in LB+bLB approach). 

\begin{figure}[htb] 
\begin{minipage}[b]{.49\linewidth}
\centering\includegraphics[width=85mm]{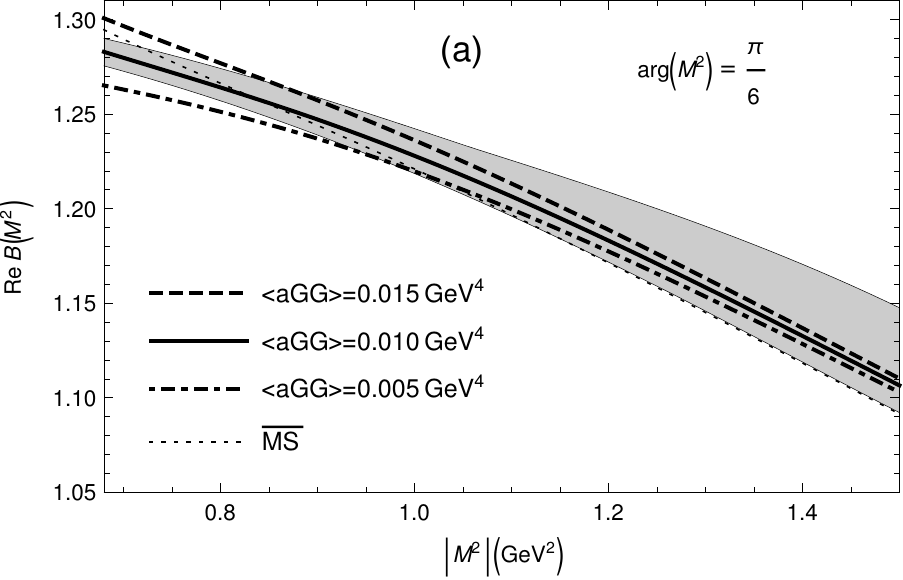}
\end{minipage}
\begin{minipage}[b]{.49\linewidth}
\centering\includegraphics[width=85mm]{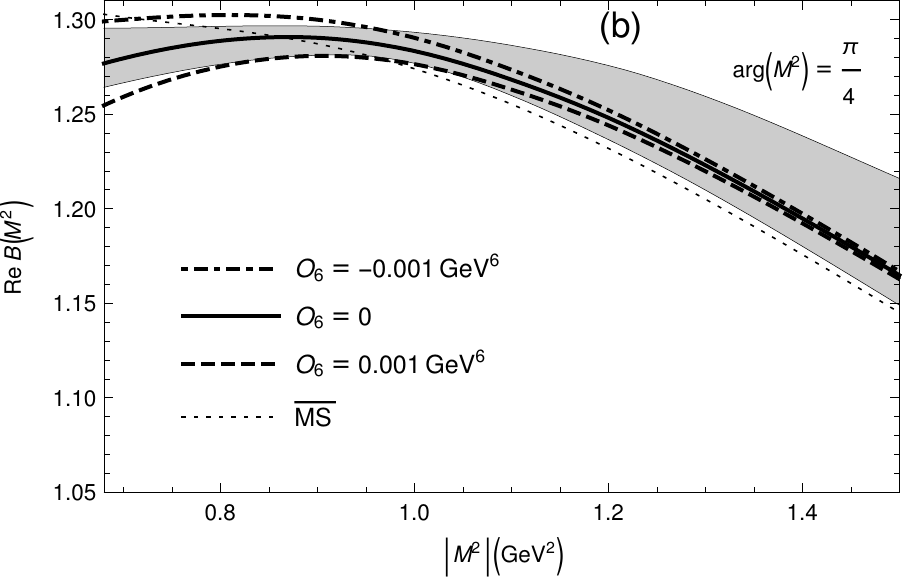}
\end{minipage}
\vspace{-0.2cm}
 \caption{\footnotesize (a) ${\rm Re} B(M^2)$ for $M^2=|M^2| \exp(i \pi/6)$; 
(b) for $M^2=|M^2| \exp(i \pi/4)$. 
The grey band represents the experimental data.
In Fig.~(a), the (EE scheme) theoretical curves correspond to $\langle a G^{\alpha}_{\mu \nu} G^{\alpha}_{\mu \nu} \rangle = (0.010 \pm 0.005) \ {\rm GeV}^4$; the $\MSbar$ curve
with $\langle a G^{\alpha}_{\mu \nu} G^{\alpha}_{\mu \nu} \rangle = 0.0059 \ {\rm GeV}^4$ is included as the dotted curve. In Fig.~(b), the (EE scheme) theoretical curves 
correspond to $\langle O^{\rm (V+A)} \rangle_6 = ( 0 \pm 0.001) \ {\rm GeV}^6$;
the $\MSbar$ curve with $\langle O^{\rm (V+A)} \rangle_6 =-1.8 \times 10^{-3} \ {\rm GeV}^6$ is included as the dotted curve.}
\label{PsiPi64}
 \end{figure}

The experimental and the theoretical results are given in 
Figs.~\ref{PsiPi64}(a),(b) for $\psi \equiv {\rm arg}(M^2) = \pi/6$, $\pi/4$, 
respectively, for the interval 
$0.68 \ {\rm GeV}^2 < |M^2| < 1.50 \ {\rm GeV}^2$. 
Comparison of the (EE scheme) theoretical curves with the experimental
bands allows us to make an ``educated guess'' estimate of the condensate values
\bes
\label{aGGO6est}
\bea
\langle a G^{\alpha}_{\mu \nu} G^{\alpha}_{\mu \nu} \rangle &=& 
(0.010 \pm 0.005) \ {\rm GeV}^4 \ ,
\label{aGGest}
\\
\langle O_6^{\rm (V+A)} \rangle &=& ( 0 \pm 0.001) \ {\rm GeV}^6 \ .
\label{O6est}
\eea
\ees
This can be compared with the values extracted by the
Borel sum rule approach in $\MSbar$ scheme [where
truncated series is taken for $d(Q^2)$] in Ref.~\cite{anOPE}:
$\langle a GG \rangle = (0.0059 \pm 0.0049) \ {\rm GeV}^4$ and 
$\langle O_6^{\rm (V+A)} \rangle = (-1.8 \pm 0.9) \times 10^{-3} \ {\rm GeV}^6$. 
It is interesting that at $\psi=\pi/4$ we extract (in the EE scheme)
the values of $\langle O_6^{\rm (V+A)} \rangle$, Eq.~(\ref{O6est}),
 which are compatible with nonnegative values. This nonnegativity
is compatible with the expectation based on the 
 vacuum saturation approximation Eq.~(\ref{aqq1}).
On the other hand, the extracted values of $\langle O_6^{\rm (V+A)} \rangle$
in the $\MSbar$ Borel sum rule approach are not compatible with
the nonnegativity Eq.~(\ref{aqq1}).

\begin{figure}[htb] 
\begin{minipage}[b]{.49\linewidth}
\centering\includegraphics[width=85mm]{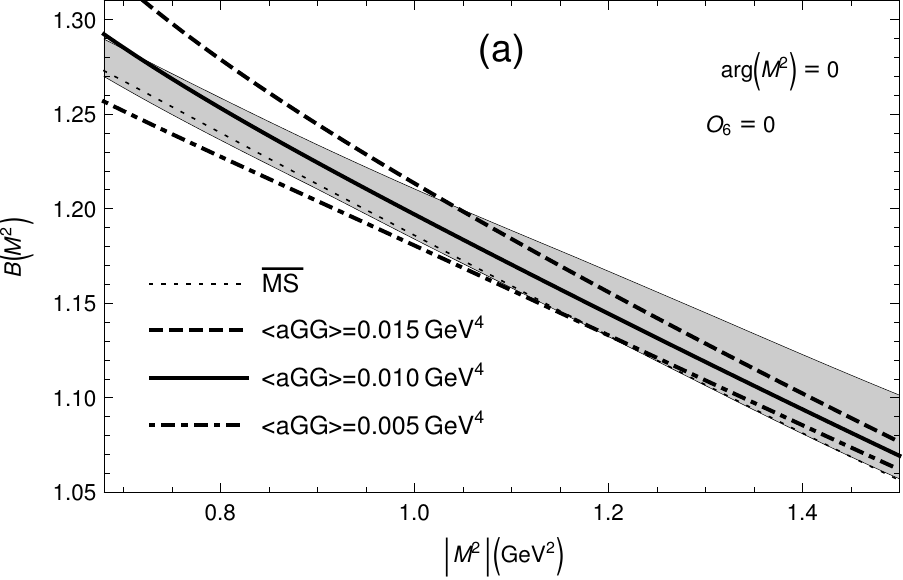}
\end{minipage}
\begin{minipage}[b]{.49\linewidth}
\centering\includegraphics[width=85mm]{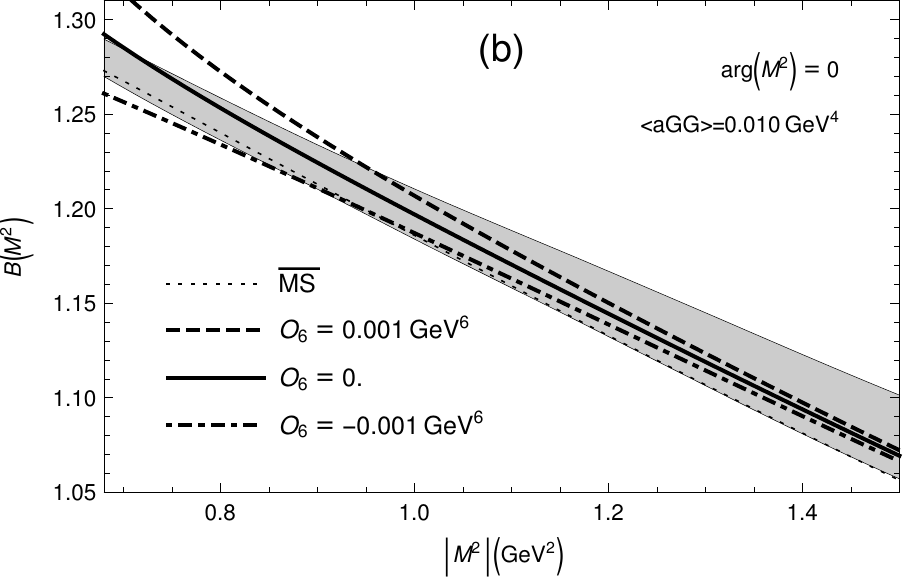}
\end{minipage}
\vspace{-0.2cm}
 \caption{\footnotesize $B(M^2)$ for $M^2=|M^2|$ (positive) scales. 
The grey band represents the experimental data.
In Fig.~(a), the (EE scheme) theoretical curves correspond to 
$\langle a G^{\alpha}_{\mu \nu} G^{\alpha}_{\mu \nu} \rangle = 
(0.010 \pm 0.005) \ {\rm GeV}^4$ and
the central value $\langle O_6^{\rm (V+A)} \rangle = 0$. 
In Fig.~(b), the (EE scheme) theoretical curves correspond to
$\langle O_6^{\rm (V+A)} \rangle = ( 0 \pm 0.001) \ {\rm GeV}^6$ 
and the central value 
$\langle a G^{\alpha}_{\mu \nu} G^{\alpha}_{\mu \nu} \rangle = 0.010 \ {\rm GeV}^4$. 
For comparison, the $\MSbar$ curve is 
included in both figures as the dotted line,
with its own central values 
$\langle a G^{\alpha}_{\mu \nu} G^{\alpha}_{\mu \nu} \rangle = 0.0059 \ {\rm GeV}^4$ 
and $\langle O_6^{\rm (V+A)} \rangle =(-1.8) \times 10^{-3} \ {\rm GeV}^6$.}
\label{Psi0}
 \end{figure}
In Figs.~\ref{Psi0}(a),(b) we present the experimental and theoretical
results for  $\psi \equiv {\rm arg}(M^2) = 0$: in Fig.~\ref{Psi0}(a)
for the choice of the obtained central value of $\langle O_6^{\rm (V+A)} \rangle$
[$=0$, Eq.~(\ref{O6est})], varying 
$\langle a G^{\alpha}_{\mu \nu} G^{\alpha}_{\mu \nu} \rangle$ in the 
obtained interval (\ref{aGGest});  in Fig.~\ref{Psi0}(b) for the
choice of the obtained central value of 
$\langle a G^{\alpha}_{\mu \nu} G^{\alpha}_{\mu \nu} \rangle$
[$=0.010 \ {\rm GeV}^4$, Eq.~(\ref{aGGest})], varying
$\langle O_6^{\rm (V+A)} \rangle$ in the obtained interval (\ref{O6est}).
Comparison with the experimental band for $\psi=0$ indicates a good agreement, 
especially for the theoretical central (full line) curve. Furthermore,
comparison with the $\MSbar$ $\psi=0$ curve (with its own central values
$0.0059  \ {\rm GeV}^4$ and $-1.8 \times 10^{-3} \ {\rm GeV}^6$) indicates
that the obtained central curve of the EE scheme is better. 

In Table \ref{tabReB}, we display various terms in the evaluation 
of the real part of the Borel transform $B(M^2)$, in the EE scheme
(analytic) pQCD, with the LB+bLB approach, for various values
of the complex scale $M^2$: $M^2 = 0.8 \exp(i \psi) \ {\rm GeV}^2$ and
$M^2 = 1.2 \exp(i \psi) \ {\rm GeV}^2$, with $\psi=\pi/6$, $\pi/4$, $0$.
These terms are based on the OPE expansion (\ref{sr3c}), where
for $D=4$ and $D=6$ condensates we take our central extracted values.
The $D=0$ contribution is given in Appendix C, Eq.~(\ref{BLBbLB}),
in conjunction with Eqs.~(\ref{BbLB2}) and (\ref{BLB3}). We denote as
``LO'' term the real part of the first term in Eq.~(\ref{BLBbLB}), i.e., 
${\rm Re}(1 - \exp(- m_{\tau}^2/M^2))$, which comes from the first (unity)
term in the expansion of the full Adler function, Eq.~(\ref{OPE2}).
\begin{table}
\caption{Various contributions to the real part ${\rm Re} B(M^2)$ of the 
Borel transform in the LB+bLB approach, in the (analytic) EE scheme pQCD,
for various complex scales: $M^2 = 0.8 \exp(i \psi) \ {\rm GeV}^2$ and
$M^2 = 1.2 \exp(i \psi) \ {\rm GeV}^2$, with $\psi=\pi/6$, $\pi/4$, $0$.
The various contributions are those indicated in Eq.~(\ref{sr3c})
and Eq.~(\ref{BLBbLB}), in conjunction with Eqs.~(\ref{BbLB2}) and (\ref{BLB3}).
The ``LO'' term is the leading term ${\rm Re}(1 - \exp(- m_{\tau}^2/M^2))$
in Eq.~(\ref{BLBbLB}). For the $D=4$ term we used the central extracted value
$\langle O_4^{\rm (V+A)} \rangle \equiv (1/6) \langle a G G \rangle =
(1/6) \times 0.010 \ {\rm GeV}^2$; the $D=6$ term is taken to be zero, in accordance
with the central extracted value $\langle O_6^{\rm (V+A)} \rangle=0$.}
\label{tabReB}  
\begin{ruledtabular}
\begin{tabular}{c c|l|cccc|c|cc}
$|M^2|$ & ${\rm arg}(M^2)$ & ${\rm Re} B$: LO & ${\rm Re} B$: LB & NLB & ${\rm N}^2{\rm LB}$ &  ${\rm N}^3{\rm LB}$ & sum  ${\rm Re} B(D=0)$ &  ${\rm Re} B(D=4)$ & ${\rm Re} B$
\\
\hline
0.8  & $\pi/6$ & 1.01284 & 0.134652 & 0.000843 & 0.080223 & 0.009826 & 1.23839 & 0.025702 & 1.26409
\\
1.2 & $\pi/6$ & 0.974163 & 0.120230 & 0.000719 & 0.067647 & 0.008934 & 1.17169 & 0.011423 & 1.18311
\\
\hline
0.8  & $\pi/4$ & 1.05762 & 0.137521 & 0.000843 & 0.081522 & 0.011219 & 1.28873 & 0 & 1.28873
\\
1.2  & $\pi/4$ & 1.04447 & 0.124725 & 0.000727 & 0.068476 & 0.009661 & 1.24806 & 0 & 1.24806
\\
\hline
0.8  & 0 & 0.980691 & 0.132679 & 0.0008434 & 0.079016 & 0.008587 & 1.20182 & 0.051404  & 1.25322
\\
1.2  & 0 & 0.928025 & 0.117452 &  0.000715 & 0.067111 &0.008301  & 1.1216 & 0.022846 & 1.14445
\end{tabular}
\end{ruledtabular}
\end{table}
From the Table we can see that the LB+bLB series has a rather good
convergence behavior. Namely, the LB term is always significantly
larger than the bLB contribution.\footnote{
The NLB contribution is very small, principally because the
NLB coefficient $T_1=1/12$ is very small, cf.~Eq.~(\ref{BbLB2}). 
This is to be compared with the ${\rm N}^2{\rm LB}$ and 
${\rm N}^3{\rm LB}$ coefficients, $T_2=103.538$ and $T_3=235.216$.}

While the theoretical curves and experimental bands
shown in Figs.~\ref{PsiPi64} can be
interpreted as representing an extraction of the 
$D=4$ and $D=6$ condensate values Eqs.~(\ref{aGGO6est}), 
the resulting theoretical curves in Figs.~\ref{Psi0} represent our
theoretical predictions for the Borel transform $B(M^2)$ 
for a continuous set of (positive) scales $M^2$ as compared to
the corresponding experimental values.

The extracted values (\ref{aGGO6est}) can be compared with the corresponding 
values obtained in the literature. The spread of these extracted values
is usually $\delta \langle a G G \rangle \approx \pm 0.005 \ {\rm GeV}^4$ 
and $\delta \langle O_6^{\rm (V+A)} \rangle \approx \pm 10^{-3} \ {\rm GeV}^6$, 
if the experimental uncertainties (the spread of the grey bands) are
considered to be the dominant source of the uncertainties.

In Ref.~\cite{anOPE}, where the Borel transform sum rules were used,
the extracted values for the gluon condensate
were $\langle a G G \rangle \approx (0.006 \pm 0.005) \ {\rm GeV}^4$
for the $\MSbar$ (truncated) and the resummed Lambert scheme pQCD and 
2$\delta$anQCD model;
$(0.010 \pm 0.006) \ {\rm GeV}^4$ for the 2$\delta$anQCD model (truncated);
$(0.012 \pm 0.005) \ {\rm GeV}^4$ for the Lambert scheme pQCD (truncated).
Here, the Lambert scheme was the scheme used for the 2$\delta$anQCD 
analytic model of Ref.~\cite{2danQCD} ($c_2 \equiv \beta_2/\beta_0 = -4.76$; $c_3=c_2^2/c_1 \approx 12.74$; etc.).\footnote{
The pQCD coupling $a(Q^2)$ in the Lambert scheme is not holomorphic,
and neither is in $\MSbar$ scheme. In comparison, in $\MSbar$ (with $N_f=3$) we have: $c_2 \approx 4.47$, 
$c_3 \approx 20.99$.}  
The truncated results were those based on the truncated series of the 
canonical Adler function $d(Q^2)$ (including $\sim a^4$);
the resummed versions were those based on a $[2/2]$-Pad\'e-related
resummation of that truncated series (cf.~Ref.~\cite{anOPE} for more details).
Further, the approximate values of the $D=6$ condensate 
$\langle O_6^{\rm (V+A)} \rangle$ extracted in Ref.~\cite{anOPE} were  
approximately $(-2 \pm 1) \times 10^{-3} \ {\rm GeV}^6$ 
for the aforementioned approaches
($\MSbar$ truncated; Lambert scheme pQCD truncated and resummed; 2$\delta$anQCD
truncated), with the exception of the resummed 2$\delta$anQCD where it was 
$(-0.5 \pm 1.1) \times 10^{-3} \ {\rm GeV}^6$.

On the other hand, earlier analyses with Borel transform sum rules,
performed in $\MSbar$ scheme, gave \cite{Geshkenbein}
$\langle a G G \rangle = (0.006 \pm 0.012) \ {\rm GeV}^4$
and  \cite{Ioffe} $(0.005 \pm 0.004) \ {\rm GeV}^4$.\footnote{
In Ref.~\cite{Ioffe}, in addition to the $V+A$ channel of the $\tau$ decay data
of ALEPH 1998 \cite{ALEPH1}, the charmonium sum rules were applied.}
In Ref.~\cite{O6sr1}, weighted finite energy sum rules and ALEPH 2005 data 
were used; the obtained values were 
$\langle a G G \rangle = (0.008 \pm 0.005)  \ {\rm GeV}^4$. In the original
work on sum rules, Ref.~\cite{Shifman1,Shifman2}, the value
$\langle a G G \rangle = 0.012  \ {\rm GeV}^4$ 
was obtained, using charmonium physics.
Application of the QCD-moment and QCD-exponential moment sum rules 
for heavy quarkonia, Refs.~\cite{O4sr1a,O4sr1b}, gave there the values 
$\langle a G G \rangle = (0.022 \pm 0.004) \ {\rm GeV}^4$
and $(0.024 \pm 0.006) \ {\rm GeV}^4$, respectively. Furthermore,
a combined fit of the $V+A$ channel $\tau$ decay data,
Ref.~\cite{DDHMZ}, extracted the value
of $\alpha_s$ and, as a byproduct, the condensate value
 $\langle a G G \rangle = (-0.015 \pm 0.003) \ {\rm GeV}^4$.

Most of the analyses in the literature give 
$\langle O_6^{\rm (V+A)} \rangle < 0$, cf.~Refs.~\cite{O6sr1,O6sr2,anOPE},
suggesting a qualitative failure of the vacuum saturation approximation
Eq.~(\ref{aqq1}), in contrast to the result (\ref{O6est}) obtained here
in the considered analytic pQCD in the EE scheme.

\section{Conclusions}
\label{sec:summ}

We constructed a perturbative mass independent beta function $\beta(a)$ for the
QCD running coupling $a(Q^2)$ ($\equiv \alpha_s(Q^2)/\pi$) at $N_f=3$ such that
the following two restrictions are fulfilled simultaneously: 
(a) the correct value of the semihadronic strangeless 
tau lepton decay ratio $r_{\tau} \approx 0.203$ is reproduced,
$r_{\tau}$ being here presently
the best measured inclusive low-energy QCD quantity with strongly
suppressed higher-twist contributions;
(b) the coupling $a(Q^2)$ has no (unphysical) Landau singularities, 
i.e., it is a holomorphic
function in the complex $Q^2$ plane  
$Q^2 \in \mathbb{C} \backslash (-\infty, -M_{\rm thr}^2]$,
with a threshold mass $M_{\rm thr} \sim 0.1$ GeV. This construction
was not straightforward, because the two mentioned conditions
tend to mutually exclude each other.
In contrast to the results of
Refs.~\cite{anpQCD1,anpQCD2}, where the growth of the coefficients $d_n$
of the spacelike physical quantities ${\cal D}(Q^2)$ due to the
scheme (beta function) was out of control already at $n=4$, we construct here
beta functions which do not lead to an explosive growth
of the coefficients $d_n$, at least up to a given chosen order $n$.
In one case (EE scheme), we even obtained a beta function which contributes 
for large $n$ to the growth of the coefficients $d_n$ 
less than the leading $b = \pm 1$ renormalon contributes.
Stated otherwise, the effects of our perturbative beta function
did not overshadow the renormalon growth of the coefficients $d_n$ and, 
at the same time, they eliminated the Landau singularities of the
running coupling and allowed the reproduction of the 
correct value of $r_{\tau}$. 
The attractiveness of the obtained holomorphic (analytic) QCD
models is that they are perturbative, i.e., beta function $\beta(a)$
is fully described by the Taylor series in powers of $a$, it has no
nonperturbative contributions such as $\exp(-{\cal K}/a(Q^2)) \sim
1/(Q^2)^{M}$, in contrast to the presently known analytic QCD models 
Refs.~\cite{ShS1,ShS2,MS,Nest1a,Nest1b,Nest1c,Nest1d,Nest2a,Nest2b,Nest2c,Webber,CV12a,CV12b,Alekseev,1danQCD,2danQCD,Shirkovmass,mes2a,mes2b,Simonov1,Simonov2,BKSKKSh1,BKSKKSh2,BKSKKSh3,BKSKKSh4,BKSKKSh5}. 
In addition, with the EE scheme (analytic) pQCD, we performed an
analysis with Borel sum rules for the $V+A$ channel of semihadronic
strangeless decays of $\tau$ lepton, and extracted reasonable
values of the corresponding condensates:
$\langle a GG \rangle = (0.010 \pm 0.005) \ {\rm GeV}^4$ and
$\langle O_6^{\rm (V+A)} \rangle = (0 \pm 0.001) \ {\rm GeV}^6$.
It remains to be seen how the presented
holomorphic pQCD models work in the evaluation of other low-momentum
inclusive observables, such as Bjorken polarized sum rule (BSR).
In contrast to the well-measured low-momentum quantity $r_{\tau}$
($V+A$ channel),
the BSR has strong chirality-conserving 
higher-twist effects at low momenta, which makes
the evaluation of this quantity even in analytic QCD models
more difficult \cite{Khandr1,Khandr2,Khandr3,Khandr4}.

\section*{Acknowledgments}
\noindent
We thank P.M.~Stevenson for very useful comments.
This work was supported in part by FONDECYT (Chile) Grant No.~1130599
and DGIP (UTFSM) internal project USM No.~11.13.12 (C.C. and G.C.),
and in part by  FONDECYT (Chile) Grant No.~1141260 (O.O.).

\appendix

\section{Leading-$\beta_0$ resummation of $r_{\tau}$ and beyond}
\label{app1}

In this Appendix, we summarize those parts of 
Appendices of Refs.~\cite{anpQCD1,anpQCD2} that are relevant in this work.

The decay ratio
of the semihadronic strangeless $\tau$ lepton decays (V+A)-channel is 
\bes
\label{Rtau}
\begin{eqnarray}
R_{\tau}(\triangle S\!=\!0)  &\equiv& 
\frac{ \Gamma (\tau^- \to \nu_{\tau} {\rm hadrons} (\gamma) )}
{ \Gamma (\tau^- \to \nu_{\tau} e^- {\overline {\nu}_e} (\gamma))}
- R_{\tau}(\triangle S\!\not=\!0)
\label{Rtaudef}
\\
& = & 
3.479  \pm 0.011 \ ,
\label{Rtauexp}
\end{eqnarray}
\ees
where the measured value given above is 
extracted from measurements by the ALEPH Collaboration \cite{ALEPH2a,ALEPH2b}
and updated in Ref.~\cite{DDHMZ}.
The QCD canonic massless quantity $r_{\tau}(\triangle S=0, m_q=0)$ 
is obtained from this quantity by removing the
non-QCD (CKM and EW) factors and contributions, and the
chirality-violating quark mass contributions
\ba
\lefteqn{
r_{\tau}(\triangle S=0, m_q=0) 
= \frac{ R_{\tau}(\triangle S=0) }
{ 3 |V_{ud}|^2 (1 + {\delta}_{{\rm EW}}) } 
} 
\nonumber\\
&&
- (1 + \delta_{\rm EW}^{\prime} ) - 
\delta r_{\tau}(\triangle S=0, m_{u,d}\not=0) \ .
\label{rtgen2}
\ea
This quantity $r_{\tau}$ is QCD-canonic in the sense that
its (leading-twist) pQCD expansion is 
$r_{\tau}(\triangle S=0, m_q=0)_{\rm pt} = a + {\cal O}(a^2)$.
The higher-twist massless contributions in this $r_{\tau}(V+A)$ are
very suppressed \cite{DDHMZ}.
Further, it is a timelike quantity, and can be expressed in terms of
the massless current-current correlation function
(V-V or A-A, both equal since massless) \cite{Tsai:1971vv}
\be
r_{\tau} = \frac{2}{\pi} \int_0^{m^2_{\tau}} \ \frac{d s}{m^2_{\tau}}
\left( 1 - \frac{s}{m^2_{\tau}} \right)^2 
\left(1 + 2 \frac{s}{m^2_{\tau}} \right) {\rm Im} \Pi(Q^2=-s) \ .
\label{rtauPi}
\ee
Using the Cauchy theorem in the complex $Q^2$ plane and
integrating by parts results in the following contour integral form
\cite{Bra1,Bra2,Bra3,Bra4,Pivovarov:1991rh,LeDiberder:1992te,Beneke:2008ad}:
\be
r_{\tau} = \frac{1}{2 \pi} \int_{-\pi}^{+ \pi}
d \phi \ (1 + e^{i \phi})^3 (1 - e^{i \phi}) \
d(Q^2=m_{\tau}^2 e^{i \phi}) \ ,
\label{rtaucont}
\ee
where $d (Q^2) = - d \Pi(Q^2)/d \ln Q^2 = a(Q^2) + {\cal O}(a^2)$ 
is the canonical massless Adler function, 
which is a spacelike QCD quantity whose
expansion in powers $a^n$ and in logarithmic derivatives $\ta_n$ is
\bes
\label{dexp}
\ba
d(Q^2)_{\rm pt} & = & a(Q^2) + \sum_{n=1}^{\infty} d_n a(Q^2)^{n+1} \ ,
\label{pt}
\\
d(Q^2)_{\rm mpt} & = & a(Q^2) + \sum_{n=1}^{\infty} {\td}_n \ta_{n+1}(Q^2) \ .
\label{mpt}
\ea
\ees
Here, the logarithmic derivatives are defined as
\be
\ta_{n+1}(\mu^2)
\equiv \frac{(-1)^n}{\beta_0^n n!}
\left( \frac{ \partial}{\partial \ln \mu^2} \right)^n a(\mu^2) \ ,
\qquad (n=1,2,\ldots) 
\label{tan}
\ee
and are related with the powers by (repeated) application of RGE
\bes
\label{tanan}
\ba
\ta_2 & = &
a^2 + c_1 a^3 + c_2 a^4 + \cdots \ ,
\label{ta2}
\\
\ta_3 & = & a^3 + \frac{5}{2} c_1 a^4 + \cdots \ ,
\qquad
\ta_4 = a^4 + \cdots \ , \qquad {\rm etc.}
\label{ta3ta4}
\ea
\ees
These relations can be recursively inverted
\bes
\label{antan}
\ba
a^2 & = & \ta_2 - c_1 \ta_3 + 
\left( \frac{5}{2} c_1^2 - c_2 \right) \ta_4 + \cdots \ ,
\label{a2}
\\
a^3 & = & \ta_3 - \frac{5}{2} c_1 \ta_4 + \cdots \ ,
\qquad  a^4 = \ta_4 + \cdots \ , \qquad {\rm etc.}
\label{a3a4}
\ea
\ees
Inserting the relations (\ref{antan}) into the power series
(\ref{pt}), we immediately obtain the coefficients ${\widetilde d}_n$
of the ``modified'' perturbation series (\ref{mpt}) in logarithmic
derivatives
\bes
\label{tdn}
\ba
\td_1 & = & d_1 \ , \qquad  \td_2 =  d_2 - c_1 d_1 \ ,
\label{td1td2}
\\
\td_3 & = & d_3 - \frac{5}{2} c_1 d_2 + 
\left(\frac{5}{2} c_1^2 - c_2 \right) d_1 \ , \qquad {\rm etc.}
\label{td3}
\ea
\ees

If the power series (\ref{pt}) is used in the contour integral
(\ref{rtaucont}) and integrated for each term separately, 
then the obtained result is called the contour improved perturbation theory
(CIPT) \cite{Pivovarov:1991rh,LeDiberder:1992te}.
However, since we consider here such schemes in which $a(Q^2)$
is holomorphic, there is another, probably better, approach 
available for the evaluation
of the contour integral (\ref{rtaucont}), which involves the so called
leading-$\beta_0$ (LB) resummation and the subsequent addition of
three other known terms. Namely, the coefficients ${\widetilde d}_n$
(and $d_n$) can be written as a power series of $N_f$, and thus as a power
series of $\beta_0$ (because $N_f=-6 \beta_0 + 33/2$)
\be
\td_n = c_{n,n} \beta_0^n + c_{n,n-1} \beta_0^{n-1} + \ldots
+ c_{n,0} \ ,
\label{tdnexpb0}
\ee
where $\td_n(LB) = c_{n,n} \beta_0^n$ is the LB part of the coefficient
$\td_n$, it is scheme independent, and is known for every $n$, 
Refs.~\cite{Broad1,Broad2}. 
It turns out that in the
series (\ref{mpt}), the LB parts can be resummed \cite{Neubert,logder1,logder2}
\bes
\label{dLB}
\bea
d^{\rm (LB)}(Q^2) &=& a(\kappa Q^2) + 
\sum_{n=1}^{\infty} c_{n,n}(\kappa) \; \beta_0^n \; {\ta}_{n+1}(\kappa Q^2) 
\label{DmptLB}
\\
 & = &   
\int_0^{\infty} \frac{dt}{t} \; F_d(t) \; a(t Q^2 e^{{\overline {\cal C}}}) \ ,
\label{DLBint2}
\eea
\ees
where ${\overline {\cal C}}=-5/3$ in $\MSbar$ scaling convention, and
the characteristic function $F_d(t)$ for Adler function is known explicitly
\cite{Neubert}\footnote{
In Ref.~\cite{Neubert} it was argued that the expression
(\ref{DLBint2}) generates the LB part of the power expansion (\ref{pt})
when $a(t Q^2 e^{{\cal C}})$ evolves according to the one-loop RGE; 
in Appendix C of Ref.~\cite{logder1} it was shown that
$a(t Q^2 e^{{\cal C}})$ can evolve according to any ($N$-)loop level
and the integral (\ref{DLBint2}) generates the LB part (\ref{DmptLB})
of the ``modified'' perturbation expansion (\ref{mpt}).}
\bes
\label{Fd}
\bea
F_d(t)_{(t<1)} &=& 
2 C_F t {\Big[} -t \ln (t) + (1+t) \ln(1+t) \ln (t) + \frac{7}{4} t 
\nonumber\\
&& 
+ (1+t)
{\rm Li}_2(-t) {\Big]} \ ,
\label{Fdtlt1}
\\
F_d(t)_{(t>1)} 
&=& 2 C_F {\Big[} \left( \frac{1}{2} + t \right) \ln (t) 
- t (1+t)  \ln (t) \ln ( 1+ 1/t )
+ \left( \frac{3}{4} +  t \right)  
\nonumber\\
&& 
+  t (1+t)
{\rm Li}_2 \left( -1/t \right) {\Big]} \ . 
\label{Fdtmt1} 
\eea
\ees
In the expressions above, $C_F=(N_c^2-1)/(2 N_c) =4/3$.
If $a(Q^2)$ [and thus ${\ta}_{n+1}(\kappa Q^2)$] have Landau singularities
at $0 < Q^2 \lesssim \ 1 {\rm GeV}^2$, the resummation
(\ref{DLBint2}) does not cure the problem of these singularities. In fact,
it makes the problem formally even worse, as the integral 
(\ref{DLBint2}) is then undefined (ambiguous) for any $Q^2 > 0$,
due to singularities in the integrand factor $a(t Q^2 e^{{\overline {\cal C}}})$
at low $t$.
On the other hand, if $a(Q^2)$ is holomorhic, no (Landau) singularities are 
encountered and the LB-integral (\ref{DLBint2}) is convergent and unambiguous.

The entire canonical Adler function
\be
d(Q^2) = d^{\rm (LB)}(Q^2) + d^{\rm (bLB)}(Q^2)
\label{dent}
\ee
consists of the LB-part (\ref{dLB}), and of the
beyond-the-leading-$\beta_0$ (bLB) contribution whose expansion is
\bea
d^{\rm (bLB)}(Q^2) & = & \sum_{n=1}^{\infty} T_n {\ta}_{n+1}(Q^2) =
\sum_{n=1}^{\infty} ( {\td}_n- c_{n,n} \beta_0^n) {\ta}_{n+1}(Q^2) \ .
\label{dbLB}
\eea

\begin{figure}[htb] 
\begin{minipage}[b]{.49\linewidth}
\centering\includegraphics[width=85mm]{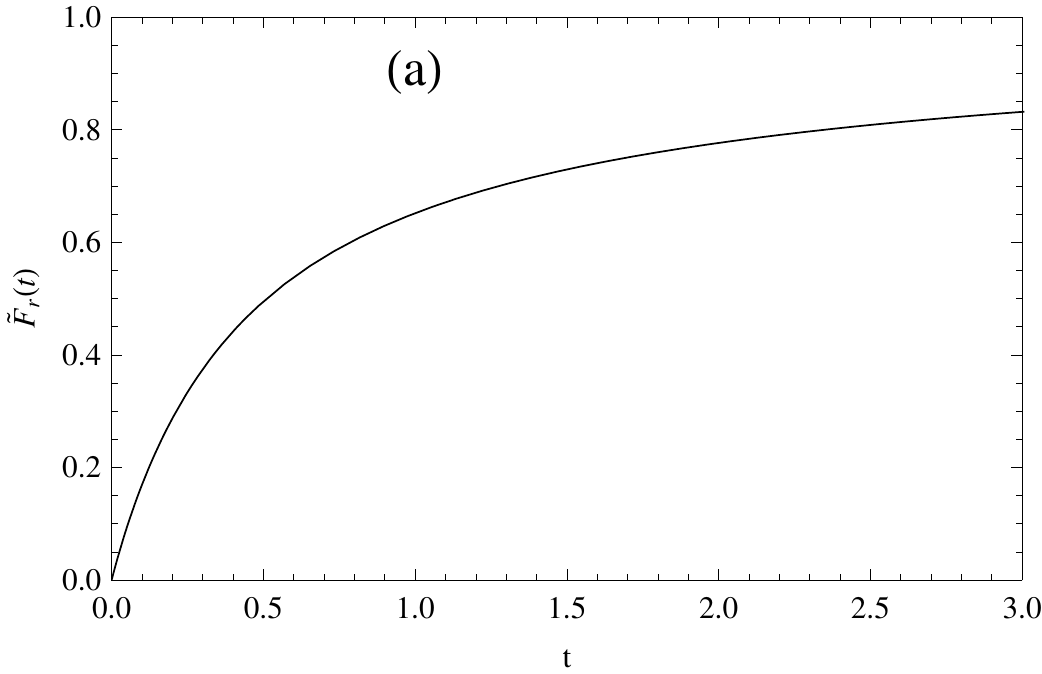}
\end{minipage}
\begin{minipage}[b]{.49\linewidth}
\centering\includegraphics[width=85mm]{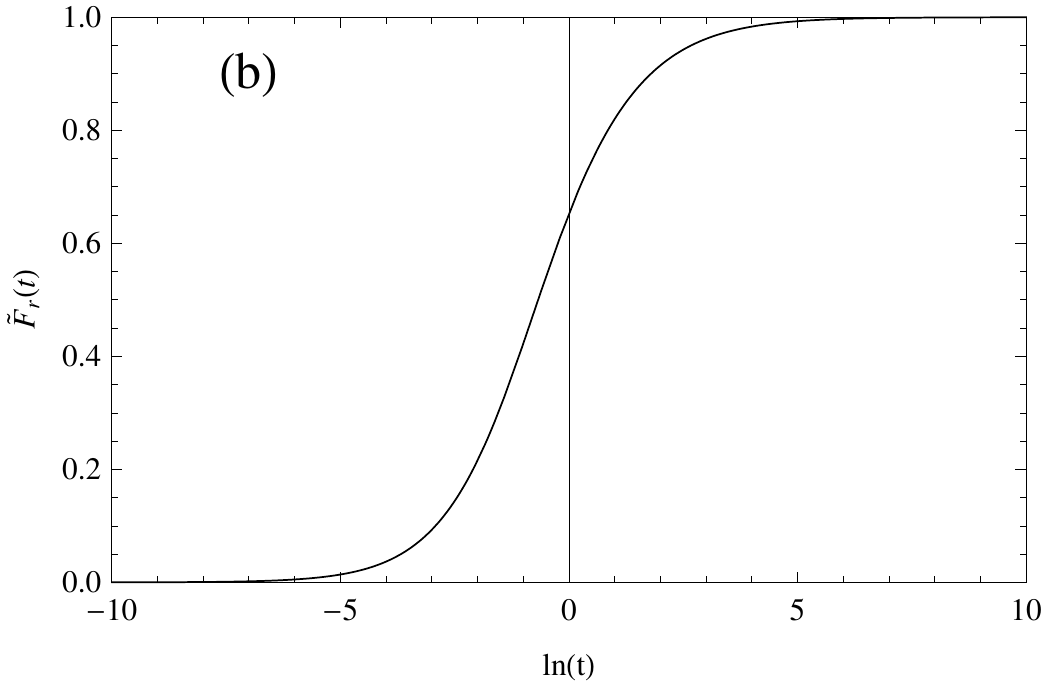}
\end{minipage}
\vspace{-0.2cm}
 \caption{\footnotesize The function ${\widetilde F}_{r}(t)$
appearing in the integral (\ref{LBrt2}) of $r_{\tau}^{\rm (LB)}$:
(a) as function of $t$; (b) as function of $\ln t$.}
\label{tFtauplz}
 \end{figure}

Insertion of the LB-integral (\ref{DLBint2}) of Adler function into the contour
integral (\ref{rtaucont}) then gives us the LB-part of $r_{\tau}$
\be
r_{\tau}^{\rm (LB)} = 
\int_0^\infty \frac{dt}{t}\: F_{r}^{\cal {M}}(t) \: 
\tlA_1 (t e^{\cal C} m_{\tau}^2) \ ,
\label{LBrt}
\ee
where the superscript ${\cal {M}}$ indicates that these
are Minkowskian (timelike) quantities;
$\tlA_1$ is the timelike coupling
\ba
\tlA_1(s) &=& \frac{1}{\pi} 
\int_s^{\infty} \frac{d \sigma}{\sigma}
{\rho}_1(\sigma) \ ;
\label{tlA1}
\ea
and the characteristic function $F_{r}^{\cal {M}}(t)$ was obtained
in Ref.~\cite{Neubert2}.\footnote{
The quantity $W_{\tau}$ of Ref.~\cite{Neubert2}
is related to $F_{r}^{\cal {M}}$ here via: $F_{r}^{\cal {M}}(t) = (t/4)W_{\tau}(t)$.}
Since $- \pi d \tlA_1(s)/d \ln s = \rho_1(s)$, integration by parts
allows us to express $r_{\tau}^{\rm (LB)}$ as an integral over the
discontinuity function $\rho_1(s) \equiv {\rm Im} \; a(-s - i \epsilon)$
\be
r_{\tau}^{\rm (LB)} = 
\frac{1}{\pi} \int_0^\infty \frac{dt}{t}\: {\widetilde F}_{r}(t) \: 
\rho_1(t e^{\cal {\overline C}} m_{\tau}^2) \ .
\label{LBrt2}
\ee 
This form is convenient here since the numerical integration of
the RGE (\ref{RGExy}) gives us the values of $\rho_1(s)$
[and not $\tlA_1(s)$],
cf.~Figs.~\ref{figrho} and \ref{figrhoEEP51}. 
The characteristic function
\be
{\widetilde F}_{r}(t) = \int_0^t \frac{dt'}{t'}\: F_{r}^{\cal {M}}(t') 
\label{tFr}
\ee
was obtained explicitly in Appendix of Ref.~\cite{anpQCD1} 
(Appendix D of Ref.~\cite{anpQCD2}); here we
only reproduce it in Figs.~\ref{tFtauplz}, for better visualization.  

However, the first three full (i.e., LB+beyond LB)
coefficients $d_1$, $d_2$ and $d_3$ ($\Rightarrow$ $\td_1$, $\td_2$, $\td_3$)
of the Adler function are known exactly \cite{d1a,d1b,d1c,d2a,d2b,d3}. 
This means that
we can add to the LB part (\ref{LBrt}) the beyond-the-leading-$\beta_0$
contributions (bLB) of order $\sim \ta_n$ ($n=1,2,3$)
\be
(r_{\tau})^{\rm (LB+bLB)} = r_{\tau}^{\rm (LB)} + 
\sum_{n=1}^{3} \ T_n I(\ta_{n+1}) \ ,
\label{rtman2}
\ee
where
\bes
\label{TnIta}
\ba
T_n &=& \td_n - \td_n^{\rm (LB)} = \td_n - c_{n,n} \beta_0^n \ ,
\label{Tn}
\\
I(\ta_{n+1}) &=&
\frac{1}{2 \pi} \int_{-\pi}^{+ \pi}
d \phi \ (1 + e^{i \phi})^3 (1 - e^{i \phi}) \
\ta_{n+1}(m_{\tau}^2 e^{i \phi}) \ ,
\label{Ita}
\ea
\ees
and $r_{\tau}^{\rm (LB)}$ is given in Eq.~(\ref{LBrt2}).
We recall that in Eq.~(\ref{Tn}) the part $\td_n^{\rm (LB)} =
c_{n,n} \beta_0^n$ is scheme independent (i.e., independent of
$c_2, c_3, \ldots$), and therefore $T_n$ has the same
scheme dependence as $\td_n$.
We consider the expression (\ref{rtman2}) in conjunction with
Eq.~(\ref{LBrt2}) as the preferred
method of evaluation, and we use it for our evaluations of $r_{\tau}$.
Implicitly, we assume that the renormalization scale 
in $d(Q^2=m_{\tau}^2 e^{i \phi})$ in
the contour integral (\ref{rtaucont}) is
$\mu^2 = Q^2$ [$=m_{\tau}^2 e^{i \phi})$]; though, other
renormalization scales could be used, e.g. 
$\mu^2= \kappa Q^2$ with $\kappa \not= 1$ ($\kappa \sim 1$).
Furthermore, we could use for the bLB contributions in Eq.~(\ref{rtman2})
the powers $a^n$, i.e., the power series of 
$d(Q^2)^{\rm (bLB)} \equiv d(Q^2) - d(Q^2)^{\rm (LB)}$ instead of the
series in logarithmic derivatives; we do not prefer this choice, because
the LB part $d(Q^2)^{\rm (LB)}$ represents a (LB-)series (\ref{DmptLB}) in
logarithmic derivatives $\ta_n$ and not in powers $a^n$.\footnote{
If LB resummation were not used, we could use the power expansion
(\ref{pt}) for the Adler function [i.e., CIPT for $r_{\tau}$ (\ref{rtaucont})]
since the considered holomorphic coupling $a(Q^2)$ is perturbative.
On the other hand, if the considered holomorphic coupling (and
beta function) were nonpertubative [$a(Q^2) \mapsto \A(Q^2)$], the use of the
expansion (\ref{mpt}) in logarithmic derivatives ($\ta_n \mapsto$)
$\tA_n$ of $\A$
(and its possible resummations) for the Adler function would be obligatory 
because otherwise the series goes out of control due to incorrect 
treatment of nonperturbative contributions \cite{logder1,logder2}.}

\section{Coefficients of perturbation expansion in a general scheme}
\label{app2}

In this Appendix we summarize the relations between the
$\MSbar$ scheme and general scheme coefficients in perturbation
expansions of physical quantities. Perturbation expansions of spacelike
observables ${\cal F}(Q^2)$ are
usually given in the literature in the $\MSbar$ scheme
\ba
{\cal F}(Q^2)_{{\rm pt} \MSbar} & = & {\overline a}^{\nu_0} + {\overline {\cal F}}_1 {\overline a}^{1+\nu_0} +
 {\overline {\cal F}}_2 {\overline a}^{2+\nu_0} + {\overline {\cal F}}_3 {\overline a}^{3+\nu_0} + \ldots \ ,
\label{ptMSbar}
\ea
where ${\overline a}$ is the coupling $a \equiv \alpha_s/\pi$ in the
$\MSbar$ renormalization scheme (${\overline c}_2,{\overline c}_3,\ldots$)
and at the canonical renormalization scale $\mu^2 = Q^2$
\be
{\overline a} \equiv a(\mu^2=Q^2; {\overline c}_2,{\overline c}_3,\ldots) \ .
\label{aMSbar}
\ee
The coupling in a different renormalization scheme ($c_2,c_3,\ldots$)
and at a general (spacelike) renormalization scale $\mu^2 = \kappa Q^2$
(where $\kappa > 0$, usually $\kappa \sim 1$)
\be
a \equiv a(\mu^2=\kappa Q^2; c_2,c_3,\ldots) 
\label{agen}
\ee
can be related to ${\overline a}$ by use of the relations (\ref{match}) 
and (\ref{RGE1})-(\ref{betaexp})
(cf.~Appendix A of Ref.~\cite{Stevenson}, and Appendix A of
Ref.~\cite{CK})
\ba
{\overline a} & = & a + a^2 \beta_0 \ln \kappa + a^3 \left[ \beta_0^2 \ln^2 \kappa + c_1 \beta_0 \ln \kappa - (c_2- {\overline c}_2) \right]
\nonumber\\
&& + a^4 \left[ \beta_0^3 \ln^3 \kappa + \frac{5}{2} c_1 \beta_0^2 \ln^2 \kappa +
c_2 \beta_0 \ln \kappa - 3 (c_2 - {\overline c}_2) \beta_0 \ln \kappa - \frac{1}{2}
(c_3 - {\overline c}_3) \right] 
\nonumber\\
&&
+ {\cal O}(a^5) \ ,
\label{abarvsa}
\ea
where the notations used are those of Eq.~(\ref{betaexp}), with
$\beta_0 = (1/4)(11 - 2 N_f/3)$ and $c_k \equiv \beta_k/\beta_0$.
Substituting the expansion (\ref{abarvsa}) into the expansion (\ref{ptMSbar}),
and performing power expansion there in powers of $a$, we obtain the 
perturbation expansion of the physical 
spacelike quantity ${\cal F}(Q^2)$ expressed in the general scheme
\ba
{\cal F}(Q^2)_{\rm pt} & = & a^{\nu_0} + {\cal F}_1 a^{1+\nu_0} +
 {\cal F}_2 a^{2+\nu_0} + {\cal F}_3 a^{3+\nu_0} + \ldots \ ,
\label{ptgen}
\ea
where the new coefficients ${\cal F}_j$ are expressed by the original
$\MSbar$ ``canonical'' coefficients  ${\overline {\cal F}}_k$ in
the following way:
\bes
\label{FvsbF}
\ba
 {\cal F}_1 & = & {\overline {\cal F}}_1 + \nu_0 \beta_0 \ln \kappa ,
\label{FvsbFa}\\
{\cal F}_2 & = & {\overline {\cal F}}_2 
+ {\overline {\cal F}}_1 (\nu_0 + 1)
\beta_0 \ln \kappa 
\nonumber\\ &&
+ \nu_0 \left[ \frac{1}{2} (\nu_0+1) \beta_0^2 \ln^2 \kappa +
c_1 \beta_0 \ln \kappa - (c_2 - {\overline c}_2) \right] ,
\label{FvsbFb}\\
{\cal F}_3 & = & {\overline {\cal F}}_3 + {\overline {\cal F}}_2 (\nu_0 + 2)
\beta_0 \ln \kappa 
\nonumber\\ &&
+ {\overline {\cal F}}_1 (\nu_0 + 1) 
\left[ \left( \frac{\nu_0}{2} +1 \right) \beta_0^2 \ln^2 \kappa +
c_1 \beta_0 \ln \kappa - (c_2 - {\overline c}_2) \right] 
\nonumber\\
&& + \nu_0 {\Big[} \frac{1}{6} (2 + 3 \nu_0 + \nu_0^2) \beta_0^3 \ln^3 \kappa +
\frac{1}{2} c_1 (3 + 2 \nu_0) \beta_0^2 \ln^2 \kappa 
\nonumber\\ &&
+ \left( c_2 - (\nu_0+2) (c_2 - {\overline c}_2) \right) \beta_0 \ln \kappa
- \frac{1}{2} (c_3 - {\overline c}_3) {\Big]}.
\label{FvsbFc}
\ea
\ees
Usually we have $\nu_0=1$, e.g., in the case of the Adler function $d(Q^2)$
which is the underlying spacelike quantity for the (timelike) quantity
$r_{\tau}$, cf.~Eqs.~(\ref{rtaucont})-(\ref{dexp}).
In general, the index $\nu_0$ may be noninteger, such as,
for example, in the case of
the underlying spacelike quantity for the (timelike) 
decay width of Higgs $\Gamma(H \to b {\bar b})$ \cite{BMS2,BMS3,GCAK}.

\section{Leading-$\beta_0$ resummation in Borel sum rules and beyond}
\label{app3}

Here we present the calculation of the $D=0$ part of the 
(theoretical) Borel transform
$B(M^2)$ of Eq.~(\ref{BD0}), using the LB+bLB approach described
in Appendix \ref{app1}. Applying the contour integration (\ref{BD0})
with the canonical Adler function $d(Q^2)$ written in the form
(\ref{dent}), we obtain
\be
B(M^2;D=0) = \left( 1 - \exp(-m_{\tau}^2/M^2) \right) 
+ B^{\rm (LB)}(M^2) + B^{\rm (bLB)}(M^2) \ ,
\label{BLBbLB}
\ee
where the bLB part is
\bes
\label{BbLB}
\bea
\lefteqn{
 B^{\rm (bLB)}(M^2)=}
\nonumber\\
&&   = \frac{1}{2 \pi} \int_{-\pi}^{\pi}
d \phi \; d^{\rm (bLB)}(Q^2\!=\!m_{\tau}^2 e^{i \phi}) 
\left[ 
\exp \left( \frac{m_{\tau}^2 e^{i \phi}}{M^2} \right) -
\exp \left( - \frac{m_{\tau}^2}{M^2} \right) \right]
\label{BbLB1}
\\
&=& \sum_{n=1}^3 T_n \frac{1}{2 \pi} \int_{-\pi}^{\pi} 
\ta_{n+1}(Q^2\!=\!m_{\tau}^2 e^{i \phi}) \left[ 
\exp \left( \frac{m_{\tau}^2 e^{i \phi}}{M^2} \right) -
\exp \left( - \frac{m_{\tau}^2}{M^2} \right) \right] ,
\label{BbLB2}
\eea
\ees
where $T_n$ are the bLB coefficients appearing in Eqs.~(\ref{dbLB})
and (\ref{Tn}). The summation over $n$ in Eq.~(\ref{BbLB2}) was truncated
at $n=3$, because only the first three coefficients $d_1, d_2, d_3$
($\Rightarrow \td_1, \td_2, \td_3$) are known exactly 
\cite{d1a,d1b,d1c,d2a,d2b,d3}.

The LB part in Eq.~(\ref{BLBbLB}) is obtained in the following way.
According to Eq.~(\ref{sr3b}) we have
\be
B^{\rm (LB)}(M^2) =  \int_0^{m_{\tau}^2} \; \frac{d \sigma}{M^2}
e^{-\sigma/M^2} \omega^{\rm (LB)}(\sigma) \ .
\label{BLB1}
\ee
According to Eq.~(\ref{om1}), we have
\bes
\label{omLB1}
\bea
\lefteqn{
 \omega^{\rm (LB)}(\sigma) =
2 \pi {\rm Im} \Pi^{\rm (LB)}(Q^2=-\sigma - i \epsilon)}
\nonumber\\
&& =
i \pi \left[ \Pi^{\rm (LB)}(Q^2=-\sigma + i \epsilon)-\Pi^{\rm (LB)}(Q^2=-\sigma - i \epsilon) \right]
\\
& = & i \pi \int_{Q^{'2}=-\sigma - i \epsilon}^{-\sigma + i \epsilon}
 d ( \ln Q^{'2} ) \frac{\partial  \Pi^{\rm (LB)}(Q^{'2})}{\partial \ln  Q^{'2} }
= \frac{1}{2 \pi i} \oint_{-\sigma - i \epsilon}^{-\sigma + i \epsilon}
 \frac{d Q^{'2}}{d Q^{'2}} d^{\rm (LB)}(Q^{'2}) .
\eea
\ees
The contour integration in the two integrals in the complex $Q^{'2}$ plane
is counterclockwise 
along a circle of radius $\sigma$. We use in the last integral for the
integrand $ d^{\rm (LB)}(Q^{'2})$ the integral expression (\ref{DLBint2}), 
and interchange the order of integration over $Q^{'2}$ and $t$. This gives
\be
 \omega^{\rm (LB)}(\sigma) =  \int_0^{\infty} \; \frac{dt}{t} F_d(t)
\tlA_1 (t \sigma e^{{\overline {\cal C}}}) \ ,
\label{omLB2}
\ee
where $F_d(t)$ is the characteristic function of the (LB) Adler function,
given in Eqs.~(\ref{Fd}), and 
\be
\tlA_1(s) \equiv \frac{1}{2 \pi i} 
\oint_{-s - i \epsilon}^{-s + i \epsilon} \; \frac{d Q^{'2}}{Q^{'2}} a(s) \ ,
\label{tlA12}
\ee
where $s>0$ and the contour integration is counterclockwise in the complex
$ Q^{'2}$ plane. It turns out that this expression is exactly equal to the
expression (\ref{tlA1}) for the timelike coupling $\tlA_1$ 
already encountered in Appendix \ref{app1} 
(see, for example, Refs.~\cite{Sh1Sh2a,Sh1Sh2b,GCAK}).
Inserting the expression (\ref{omLB2}) into the Borel integral (\ref{BLB1})
then gives, upon the substitution $\tau=t \sigma/m_{\tau}^2$
and interchanging the order of integration
\be
B^{\rm (LB)}(M^2) = \int_0^{\infty} \; \frac{d \tau}{\tau} F_B(\tau;M^2)
\tlA_1 (\tau m_{\tau}^2 e^{{\overline {\cal C}}}) \ ,
\label{BLB2a}
\ee
where
\be
F_B(\tau;M^2) = \frac{m_{\tau}^2}{M^2} \int_0^1 dx \exp \left(-
\frac{m_{\tau}^2}{M^2} x \right) F_d(\tau/x) \ .
\label{BLB2b}
\ee
The timelike coupling $\tlA_1(s)$ is, according to Eq.~(\ref{tlA1}), an
integral over $\sigma$ of the discontinuity function 
$\rho_1(\sigma) = {\rm Im} a(Q^2=-\sigma - i \epsilon)$ ($\sigma>0$).
This discontinuity function is a result of the numerical
integration of the RGEs (\ref{RGExy}) in the $Q^2$ complex plane,
cf.~Figs.~\ref{figrhoEEP51}. In the evaluation of
$B^{\rm (LB)}(M^2)$ we would like to 
avoid an additional integration over $\sigma$
involving $\rho_1(\sigma)$, Eq.~(\ref{tlA1}). Therefore, the trick is
to apply in (\ref{BLB2a}) integration by parts in the integral over $\tau$,
and use the identity
\be
\frac{d}{d \ln \sigma} \tlA_1(\sigma) = - \frac{1}{\pi} \rho_1(\sigma) 
\label{dtlA1}
\ee
which is a direct consequence of the identity (\ref{tlA1}). Then the LB part
of the (theoretical) Borel transform, $B^{\rm (LB)}(M^2)$, can be expressed
in the following more convenient form involving $\rho_1$
[instead of $\tlA_1$]:
\be
B^{\rm (LB)}(M^2) = \frac{1}{\pi} \int_0^{\infty} \; \frac{d t}{t} 
\tF_B(t;M^2) \rho_1(t m_{\tau}^2 e^{{\overline {\cal C}}}) \ ,
\label{BLB3}
\ee
where the function $\tF_B(t;M^2)$, 
which can be called the characteristic function of
the (LB) Borel transform  $B^{\rm (LB)}(M^2)$, is
\bes
\label{tFb1}
\bea
\tF_B(t;M^2) & = &
\int_0^{t} \; \frac{d t^{'}}{t^{'}} F_B(t^{'};M^2)
\label{tFB1a}
\\
& = & \frac{m_{\tau}^2}{M^2} 
\int_0^1 dx \exp \left( -\frac{m_{\tau}^2}{M^2} x \right)
\int_{0}^{t/x} \frac{d \tau}{\tau} F_d(\tau) \ .
\label{tFB1b}
\eea
\ees
Using the expression (\ref{Fd}) for $F_d(t)$, the integration over
$\tau$ in Eq.~(\ref{tFB1b}) can be performed explicitly, and we obtain
\bes
\label{tFB2}
\bea
\tF_B(t;M^2)_{(t<1)} & = & \frac{1}{6} C_F 
{\Bigg[} (1\! - \! e^{- {\cal K}} ) (21  \! -  \! 2 \pi^2) + 
{\cal K} \int_0^t dx e^{-{\cal K}x} f_1(t/x) 
\nonumber\\ &&
+
{\cal K} \int_t^1 dx e^{-{\cal K}x} f_2(t/x) {\Bigg]},
\label{tFB2a}
\\
\tF_B(t;M^2)_{(t>1)} & = & \frac{1}{6} C_F 
\left[ (1 \! - \! e^{- {\cal K}} ) (21 \! - \! 2 \pi^2) + 
{\cal K} \int_0^1 dx e^{-{\cal K}x} f_1(t/x) \right],
\label{tFB2b}
 \eea
\ees
where ${\cal K} \equiv m_{\tau}^2/M^2$, and $f_1$ and $f_2$ are the
following functions:
\bes
\label{f1f2}
\bea
f_1(u) & = & [ -6 + 2 \pi^2 + 6 u + 3 \left( 2 + 2 u - 2 (1+u)^2 \ln(1+u) \right)
\ln u 
\nonumber\\ &&
+ 6 (1+u)^2 {\rm Li}_2(-1/u) ] \ ,
\label{f1}
\\
f_2(u) & = & [ -21 + 2 \pi^2 + 6 u +15 u^2 + 3 \left( -u (2 + 3 u) + 2 (1+u)^2 \ln(1 + u) \right) \ln u 
\nonumber\\ &&
+ 6 (1+u)^2 {\rm Li}_2(-u) ] \ .
\label{f2}
\eea
\ees
In practice, we expanded the integrand $f_1(t/x)$ in powers of $(x/t)$
[up to $(x/t)^{10}$] and the integrand $f_2(t/x)$ in powers of $(t/x)$
[up to $(t/x)^{10}$], and performed the integrations over $x$ explicitly
term by term \cite{Math}. 
This gave us the values of the characteristic function
$\tF_B(t;M^2)$ with high precision.

\begin{figure}[htb] 
\begin{minipage}[b]{.49\linewidth}
\centering\includegraphics[width=85mm]{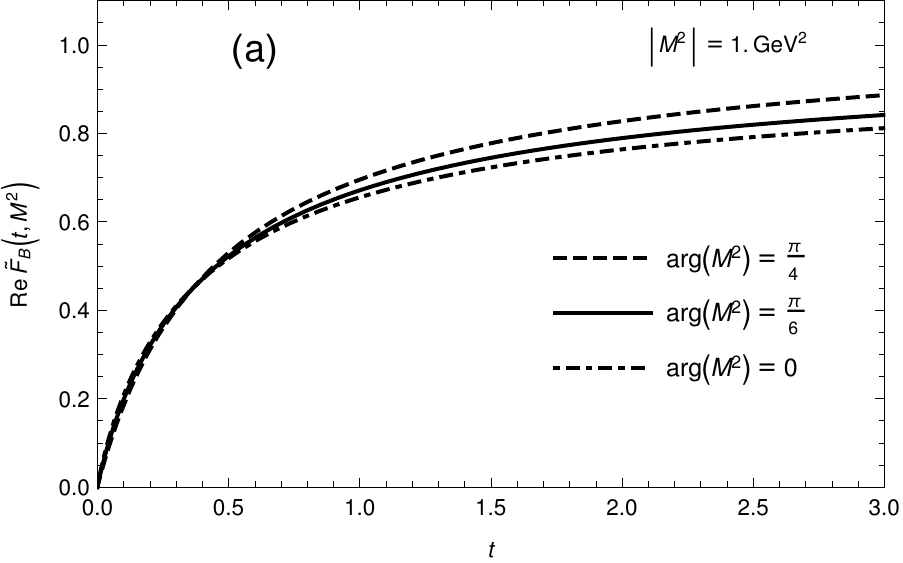}
\end{minipage}
\begin{minipage}[b]{.49\linewidth}
\centering\includegraphics[width=85mm]{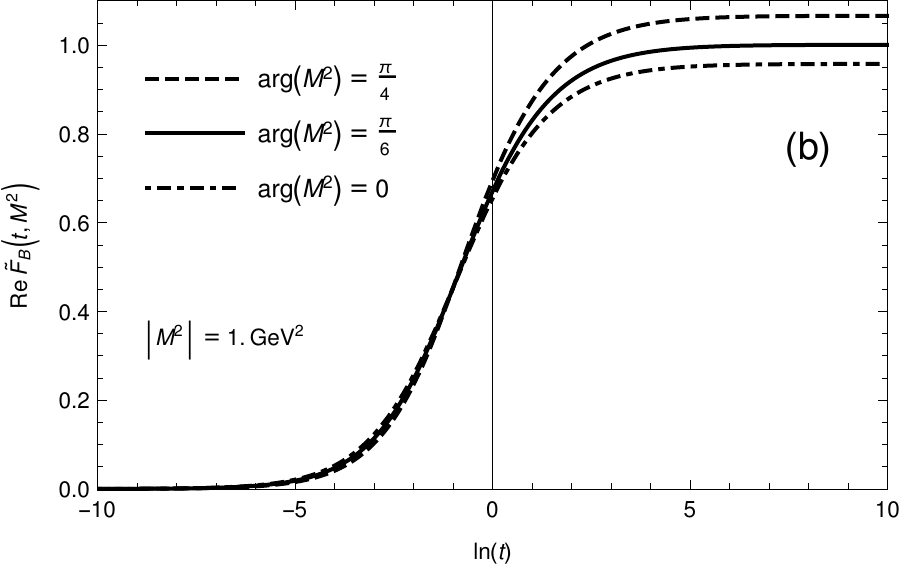}
\end{minipage}
\vspace{-0.2cm}
 \caption{\footnotesize The real part  
of the characteristic function $\tF_B(t;M^2)$ of Eq.~(\ref{tFB2}):
(a) as function of $t$; (b) as function of $\ln t$. Three different scales
$M^2$ are chosen: $|M^2|= \exp( i \pi/4),\exp( i \pi/6), 1$ (in ${\rm GeV}^2$).}
\label{tFB}
 \end{figure}
In Figs.~\ref{tFB}(a),(b) we present the real part of the
characteristic function, ${\rm Re} \tF_B(t;M^2)$,
as a function of $t$ and $\ln(t)$, for  $|M^2| = 1 \ {\rm GeV}^2$
and three choices of the arguments 
$\psi \equiv {\rm arg}(M^2) = \pi/4, \pi/6, 0$.

\end{document}